\documentclass{llncs}
\usepackage{graphicx}
\usepackage{booktabs}
\usepackage{url}
\usepackage[font=small, compatibility=false, tableposition=top]{caption}
\usepackage[pdfborder={0 0 0}]{hyperref}
\usepackage{subfig}
\usepackage{xspace}
\usepackage{amssymb}
\usepackage{amsmath}
\usepackage[mathcal]{euscript}
\usepackage[nolist,nohyperlinks]{acronym}
\usepackage[ruled,vlined,linesnumberedhidden]{algorithm2e}
\usepackage{todonotes}
\usepackage{multirow}
\usepackage{sidecap}
\usepackage{hyphenat}

\DeclareMathOperator*{\argmin}{arg\,min}

\newcommand{\pc}{{\footnotesize \,\%} }

\newcommand{\ta}{\kern0.2em}

\SetArgSty{textrm}

\SetCommentSty{commentsty}
\SetKw{Each}{each}
\ifdefined\SetAlCapSty
  \SetAlCapSty{textbf}
\fi
\ifdefined\SetAlgoCaptionSeparator
  \SetAlgoCaptionSeparator{.}
\fi
\ifdefined\DontPrintSemicolon
  \DontPrintSemicolon
\fi
\SetKw{Each}{each}

\ifdefined\algorithmautorefname
  \renewcommand{\algorithmautorefname}{Alg.\@}
\else
  
\fi

\newcommand{\ie}{{i.\,e.,\@}\xspace}
\newcommand{\eg}{{e.\,g.,\@}\xspace}
\newcommand{\etal}{{et al.\@}\xspace}
\newcommand{\cf}{{cf.\@}\xspace}

\begin{acronym}
  \acro{aca}[ACA]{Adaptive Constrained Alignment}
  \acro{ascet}[ASCET]{Advanced Simulation and Control Engineering Tool}
  \acro{atl}[ATL]{Atlas Transformation Language}
  \acro{bfs}[BFS]{breadth first search}
  \acro{cau}[CAU]{Christian-Albrechts-Universit\"at zu Kiel}
  \acro{chess}[CHESS]{Center for Hybrid and Embedded Software Systems}
  \acro{codaflow}[CoDaFlow]{Constrained Data Flow}
  \acro{cola}[CoLa]{Constrained Layout}
  \acro{dag}[DAG]{directed acyclic graph}
  \acro{dfd}[DFD]{Data Flow Diagram}
  \acro{dfs}[DFS]{depth first search}
  \acro{dsl}[DSL]{Domain Specific Language}
  \acro{dsml}[DSML]{Domain Specific Modeling Language}
  \acro{ecu}[ECU]{electronic control unit}
  \acro{eecs}[EECS]{Electrical Engineering and Computer Sciences}
  \acro{ehandbook}[EHANDBOOK]{}
  \acro{emf}[EMF]{Eclipse Modeling Framework}
  \acro{etas}[ETAS]{Engineering Tools, Application and Services}
  \acro{fsm}[FSM]{finite state machine}
  \acro{gd}[GD]{graph drawing}
  \acro{gef}[GEF]{Graphical Editing Framework}
  \acro{glmm}[GLMM]{graph layout meta model}
  \acro{gmf}[GMF]{Graphical Modeling Framework}
  \acro{graphml}[GraphML]{Graph Markup Language}
  \acro{gui}[GUI]{Graphical User Interface}
  \acro{gxl}[GXL]{Graph eXchange Language}
  \acro{hmi}[HMI]{human-machine interface}
  \acro{ide}[IDE]{integrated development environment}
  \acro{ieee}[IEEE]{Institute of Electrical and Electronics Engineers}
  \acro{ilog}[ILOG]{Intelligence Logiciel}
  \acro{ilp}[ILP]{integer linear program}
  \acro{jni}[JNI]{Java Native Interface}
  \acro{json}[JSON]{Java Script Object Notation}
  \acro{jvm}[JVM]{Java Virtual Machine}
  \acro{lncs}[LNCS]{Lecture Notes in Computer Science}
  \acro{m2m}[M2M]{Model to Model}
  \acro{m2t}[M2T]{Model to Text}
  \acro{mde}[MDE]{model-driven engineering}
  \acro{mdsd}[MDSD]{model-driven software development}
  \acro{mic}[MIC]{model-integrated computing}
  \acro{moc}[MoC]{model of computation}
  \acro{mof}[MOF]{Meta Object Facility}
  \acro{mvc}[MVC]{model-view-controller}
  \acro{kaom}[KAOM]{\acs{kieler} Actor Oriented Modeling}
  \acro{kcss}[KCSS]{Kiel Computer Science Series}
  \acro{keg}[KEG]{\acs{kieler} Editor for Graphs}
  \acro{kiel}[KIEL]{Kiel Integrated Environment for Layout}
  \acro{kieler}[KIELER]{Kiel Integrated Environment for Layout Eclipse Rich Client}
  \acro{kiem}[KIEM]{\acs{kieler} Execution Manager}
  \acro{kiml}[KIML]{\acs{kieler} Infrastructure for Meta Layout}
  \acro{klay}[KLay]{\acs{kieler} Layouters}
  \acro{klaylay}[KLay Layered]{\acs{klay}\allowbreak Layered}
  \acro{klodd}[KLoDD]{\acs{kieler} Layout of Dataflow Diagrams}
  \acro{ocl}[OCL]{Object Constraint Language}
  \acro{ogdf}[OGDF]{Open Graph Drawing Framework}
  \acro{omg}[OMG]{Object Management Group}
  \acro{qvt}[QVT]{Query-View-Transformations}
  \acro{rca}[RCA]{Rich Client Application}
  \acro{rcp}[RCP]{Rich Client Platform}
  \acro{sbgn}[SBGN]{Systems Biology Graphical Notation}
  \acro{scade}[SCADE]{Safety Critical Application Development Environment}
  \acro{ssm}[SSM]{Safe State Machines}
  \acro{sud}[SUD]{system under development}
  \acro{tmf}[TMF]{Textual Modeling Framework}
  \acro{tsm}[TSM]{topology-shape-metrics}
  \acro{ui}[UI]{user interface}
  \acro{uml}[UML]{Unified Modeling Language}
  \acro{upl}[UPL]{Upward Planarization Layout}
  \acro{vlsi}[VLSI]{very large scale integration}
  \acro{wysiwyg}[WYSIWYG]{What-You-See-Is-What-You-Get}
  \acro{xgmml}[XGMML]{Extensible Graph Markup and Modeling Language}
  \acro{xml}[XML]{Extensible Markup Language}
\end{acronym}

\pdfoutput=1
\newcommand\blfootnote[1]{%
  \begingroup
  \renewcommand\thefootnote{}\footnote{#1}%
  \addtocounter{footnote}{-1}%
  \endgroup
}

\begin{document}
\mainmatter

\title{Stress-Minimizing Orthogonal Layout\\ of Data Flow Diagrams with Ports}

\author{Ulf R\"uegg\inst{1} \and Steve Kieffer\inst{2}\and \\
  Tim Dwyer\inst{2} \and Kim Marriott\inst{2} \and Michael Wybrow\inst{2}}

\institute{
  Department of Computer Science, Kiel University, Kiel, Germany\\
  \email{uru@informatik.uni-kiel.de}
  \and 
  Faculty of Information Technology, Monash University,
  NICTA Victoria, Australia\\
  \email{\{Steve.Kieffer,Tim.Dwyer,Kim.Marriott,Michael.Wybrow\}@monash.edu}
}

\maketitle

\setcounter{footnote}{0}

\begin{abstract}
We present a fundamentally different approach to orthogonal layout of
data flow diagrams with ports. This is based on extending constrained 
stress majorization to cater for ports and flow layout.
Because we are minimizing stress we are able to better display global 
structure, as measured by
several criteria such as stress,
edge-length variance, and aspect ratio.
Compared to the layered approach, our layouts tend to
exhibit symmetries, and eliminate inter-layer whitespace, making the
diagrams more compact.
\end{abstract}

\section{Introduction}
\label{lab:introduction}
\begin{figure}[tb]
  \centering
  \subfloat[Layout with layer-based algorithm \acs{klaylay} by Schulze~\etal]{
    \includegraphics[scale=.215]{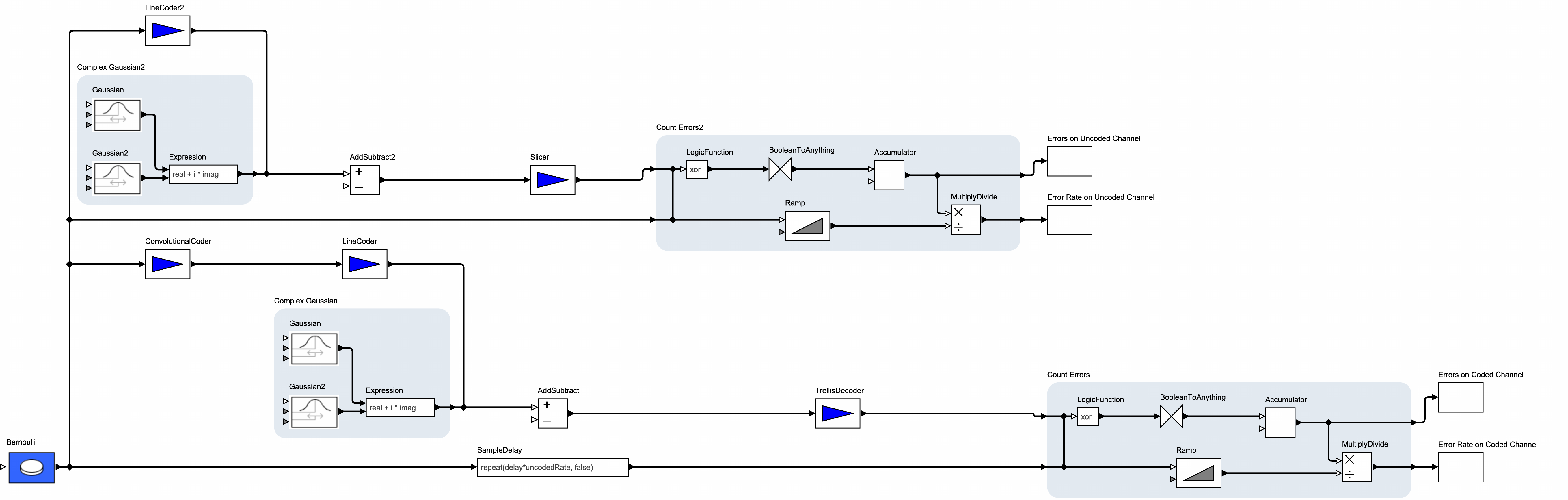} 
    \label{fig:example_klay}
  }\\
  \subfloat[Layout with the \acs{codaflow} algorithm presented here.]{
    \hspace{1em}
   \includegraphics[scale=.215]{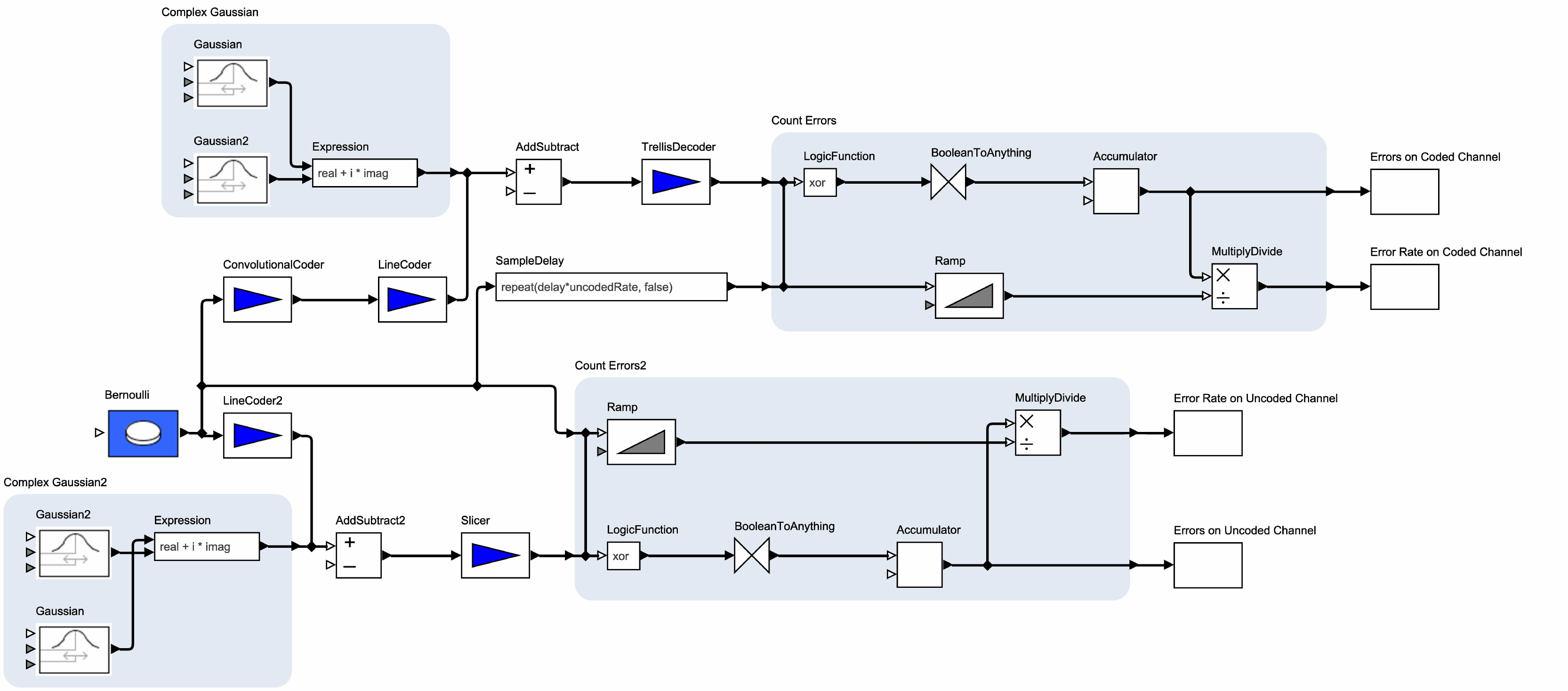} 
    \hspace{1em}
   \label{fig:example_cola}
  }
  \caption{
    Two layouts of the same diagram. The result of our method, shown
    in (b), has less stress, lower edge length variance, less area, and better
    aspect ratio. 
  }
  \label{fig:example}
\end{figure}

Actor-oriented data flow diagrams are commonly used to model movement of data between components 
in complex hardware and software systems~\cite{LeeNW03}. 
They are provided in many widely used modelling tools including 
LabVIEW (National Instruments Corporation), 
Simulink (The MathWorks, Inc.), \acs{ehandbook} (ETAS), \acs{scade} (Esterel Technologies),
and Ptolemy (UC Berkeley). Complex systems are modelled graphically by
composing \emph{actors}, \ie reusable block diagrams representing
well-defined pieces of functionality. Actors can be \emph{nested}---\ie
composed of other actors---or \emph{atomic}. \autoref{fig:example_klay}
shows an example of a data flow diagram with four nested actors. 
Data flow is shown by directed edges from the source port where the data is constructed to the 
target port where the data is consumed. By convention the edges 
are drawn orthogonally and the ports are fixed in position 
on the actors' boundaries. 
Automatic layout of data flow diagrams is important: 
Klauske and Dziobek~\cite{KlauskeD10} found that without
automatic layout about 30\pc of a
modeller's time is spent manually arranging elements.
\blfootnote{A version of this paper has been accepted for publication in Graph Drawing 2014.
The final publication will be available at \url{link.springer.com}.}

Current approaches to automatic layout of data flow diagram  are modifications of 
the well-known Sugiyama layer-based layout algorithm~\cite{SugiyamaTT81} 
extended to handle ports and orthogonal edges.
In particular Schulze~\etal~\cite{SchulzeSvH14} have spent many years 
developing specialised layout algorithms that are used, for instance, 
in the \acs{ehandbook} and Ptolemy tools.
However, their approach has a number of drawbacks. 
First, it employs a strict layering which may result in 
layouts with poor aspect ratio and poor compactness, especially when large 
nodes are present. Furthermore, the diagrams often have long edges and 
the underlying structure and symmetries may not be revealed. 
A second problem with the approach of Schulze~\etal is that 
it uses a recursive bottom-up strategy to compute a layout for 
nested actors independent of the context in which they appear.

This paper presents a fundamentally different approach to the layout 
of actor-oriented data flow diagrams designed to overcome these problems. 
A comparison of our new approach with standard layer-based
algorithm \acs{klaylay} is shown in \autoref{fig:example}. 
Our starting point is \emph{constrained stress majorization}~\cite{DwyerKM06b}.
Minimizing stress has been shown to improve readability 
by giving a better understanding of important graph structure
such as cliques, chains and cut nodes~\cite{DwyerLFQ+09}. 
However, stress-minimization typically results in a 
quite ``organic'' look with nodes placed freely 
in the plane that is quite different to the very ``schematic'' arrangement 
involving orthogonal edges, a left-to-right ``flow'' of 
directed edges, and precise alignment of node ports that practitioners prefer.

The main technical contribution of this paper is to extend constrained stress 
majorization to handle the layout conventions of data flow diagrams.  
In particular we: (1) augment the \emph{$P$-stress}~\cite{DwyerMW09a} model to
handle ports that are constrained to node boundaries but are either
allowed to float subject to ordering constraints or else are fixed to a given
node boundary side, and (2) extend \acf{aca}~\cite{KiefferDM+13} for achieving grid-like 
layout to handle directed edges, orthogonal routing,
ports, and widely varying node dimensions.

An empirical evaluation of the new approach (\autoref{sec:discussion}) shows 
it produces layouts of comparable quality to the method of Schulze~\etal 
but with a different trade off between aesthetic criteria. The layouts have more 
uniform edge length, better aspect ratio, and are more compact but
have slightly more edge crossings and bends.  
Furthermore, our method is more flexible and requires far less implementation effort. 
The Schulze~\etal approach took a team of developers and researchers 
several years to implement by extensively augmenting the Sugiyama method. 
While their infrastructure allows a flexible configuration of the 
existing functionality~\cite{SchulzeSvH14}, 
it is very restrictive and brittle when it comes 
to extensions that affect multiple phases of the algorithm.
The method described in this paper took about two months to implement and is also more 
extendible since it is built on modular components with well-defined work flows 
and no dependencies on each other.

\paragraph{Related Work.} 
The most closely related work is the series of papers 
by Schulze~\etal that show how to extend 
the layer-based approach to handle 
the layout requirements of data flow diagrams~\cite{SchulzeSvH14,SpoenemannFvH+09}. 
Their work presents several improvements over previous methods to reduce
edge bend points and crossings in the presence of ports.
While the five main phases (classically three) of 
the layer-based approach are already complex, they introduce between 
10 and 20 \emph{intermediate processes}
in order to address additional requirements.
The authors admit that their approach faces problems with 
unnecessary crossings of inter-hierarchy edges as they layout 
compound graphs bottom-up, \ie processing the most nested 
actor diagrams first. 
Related work in the context of the layer-based approach has been 
studied thoroughly in~\cite{SchulzeSvH14,SpoenemannFvH+09}. 
Chimani~\etal present methods to consider ports and their constraints
during crossing minimization within the \emph{upward planarization}
approach~\cite{ChimaniGM+10}. 
While the number of crossings is significantly reduced, the approach
eventually induces a layering, suffering from the same issues as above.
There is no evaluation with real-world examples. 
Techniques from the area of \acs{vlsi} design and 
other approaches that specifically target
compound graphs have been discussed before
and found to be insufficient to fulfil the layout requirements for 
data flow diagrams~\cite{SpoenemannFvH+09}, especially
due to lacking support for different port constraints.

\section{CoDaFlow --- The Algorithm}
\label{sec:algorithm}

\begin{figure}[tb]
  \centering
  \subfloat[After Node Positioning]{
    \includegraphics[width=.3\textwidth]{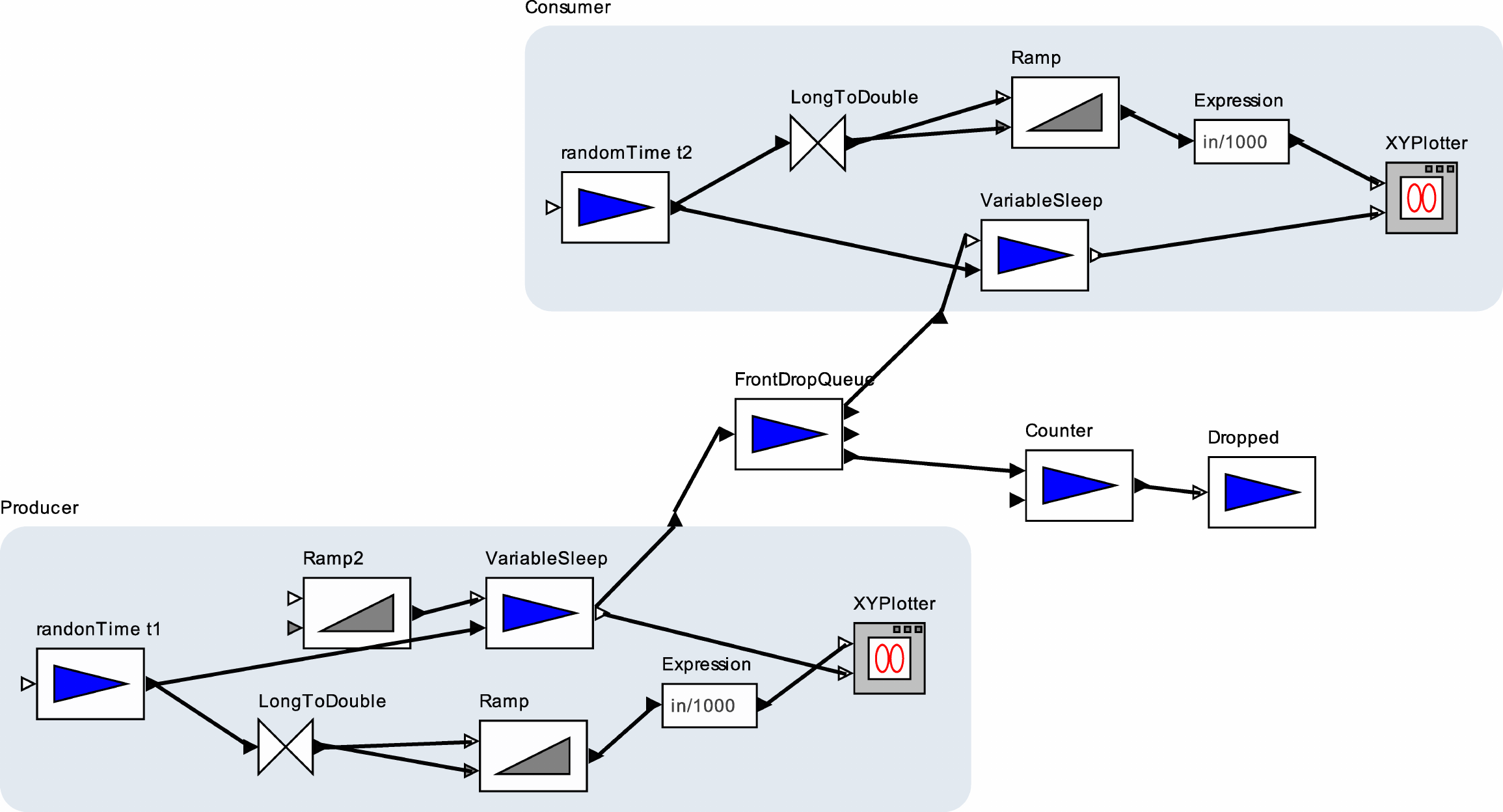} 
  }
  \subfloat[After Node Alignment]{
    \includegraphics[width=.3\textwidth]{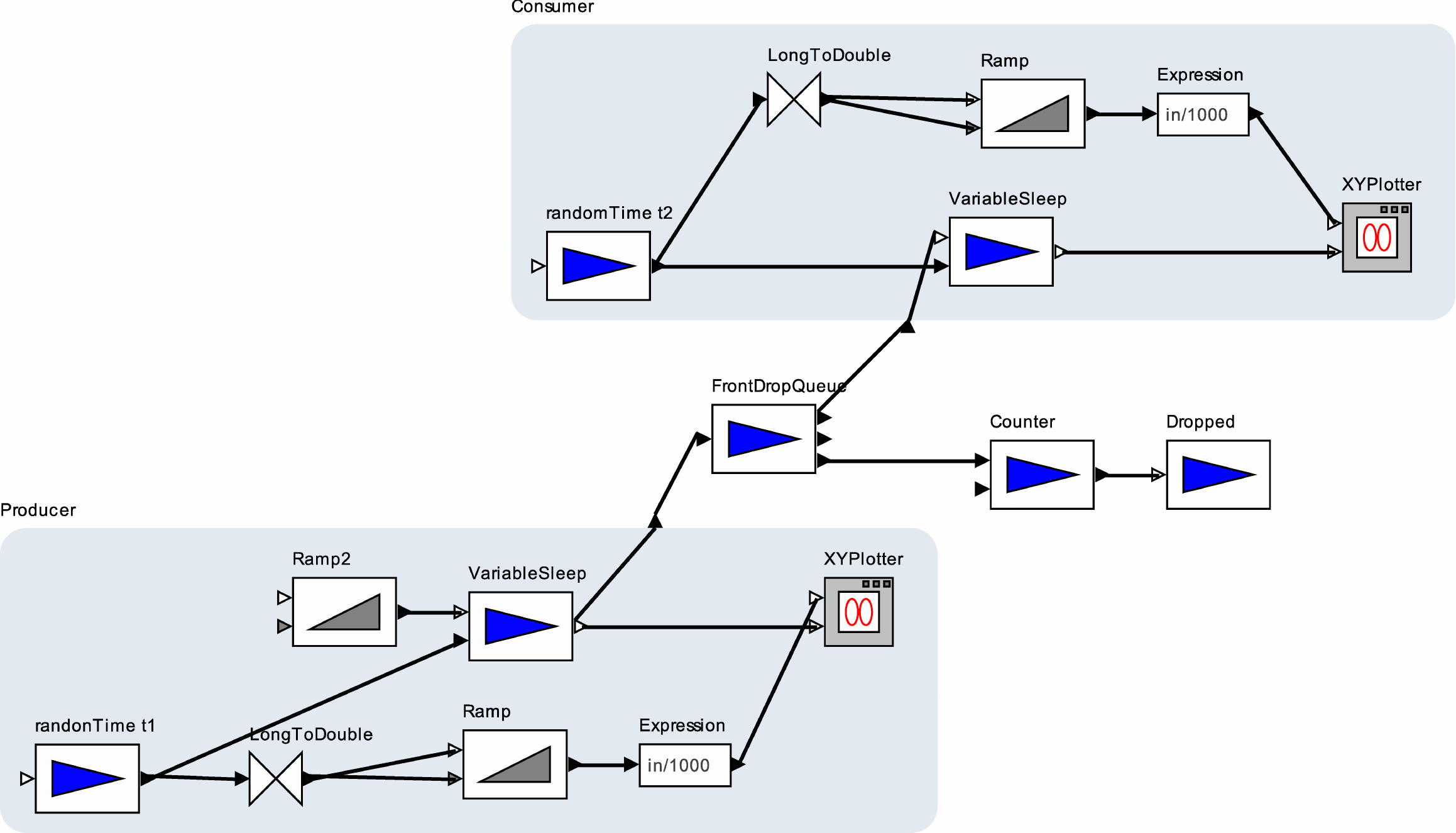} 
  }
  \subfloat[After Edge Routing]{
    \includegraphics[width=.3\textwidth]{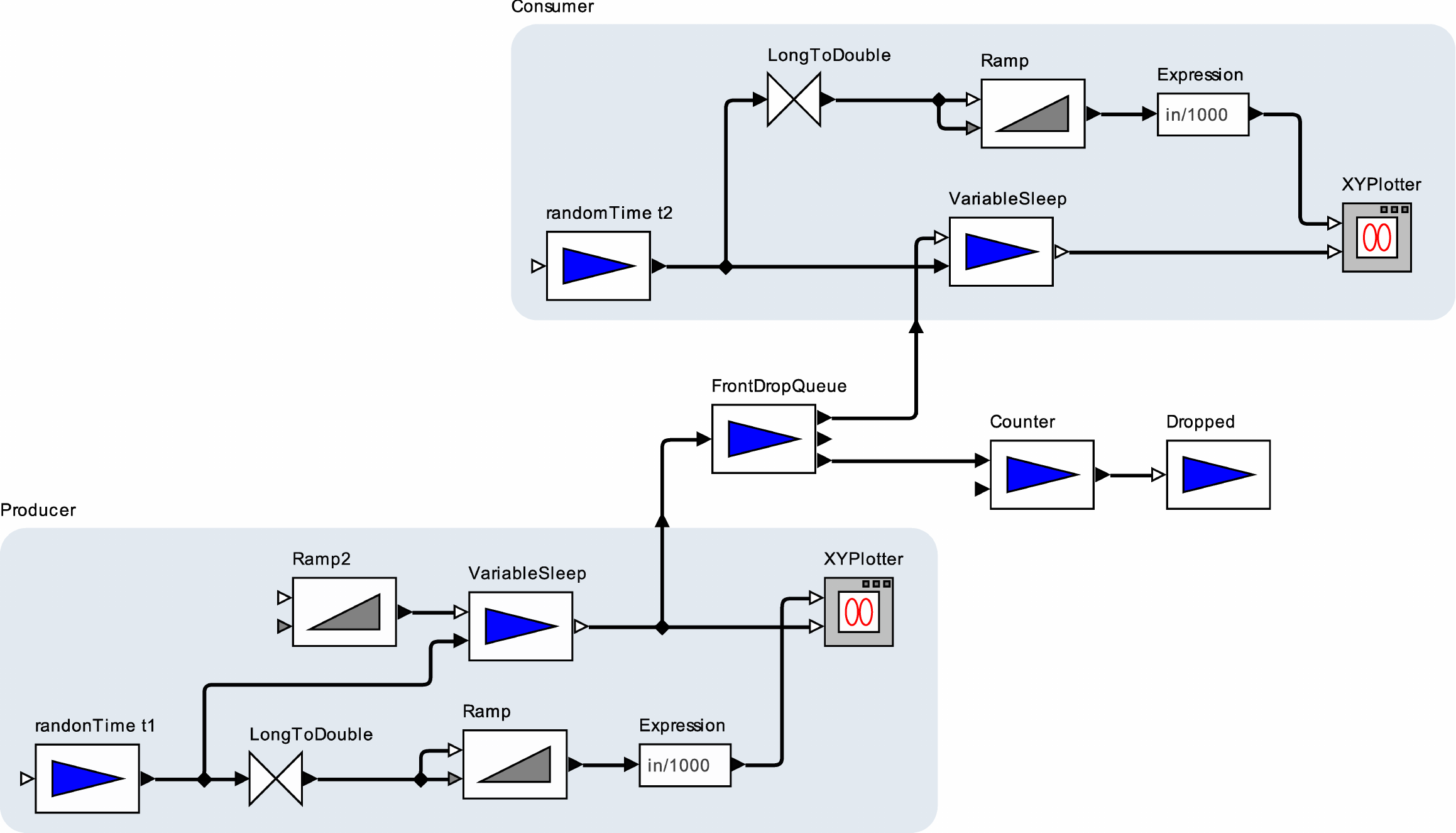} 
  }

  \caption{
    The results of pipeline stages (1), (2), (3) are shown in
    (a), (b), (c), respectively.
  }
  \label{fig:pipeline}
\end{figure}

Data flow diagrams can be modelled as directed graphs $G = (V,E,P,\pi)$
where \emph{nodes} or \emph{vertices} $v \in V$ are connected by 
\emph{edges} $e \in E \subseteq P \times P$ through \emph{ports} 
$p \in P$---certain positions on a node's perimeter---and
$\pi : P \rightarrow V$ maps each port $p$ to the \emph{parent}
node $\pi(p)$ to which it belongs.
An edge $e = (p_1, p_2)$ is directed,
\emph{outgoing} from port $p_1$ and \emph{incoming} to $p_2$.
A \emph{hyperedge} is a set of edges where every pair of edges
shares a common port.

To better show flow it is preferable for sources of edges to be to 
the left of their targets and by convention edges are routed
in an \emph{orthogonal} fashion. Ports can---depending on the
application---be restricted by certain constraints, \eg all ports with
incoming edges should be placed on the left border of the node.
Sp\"onemann~\etal define five types of \emph{port
constraints}~\cite{SpoenemannFvH+09}, ranging from 
ports being free to float arbitrarily on a node's perimeter, to
ports having well-defined positions relative
to nodes. Nodes that contain nested diagrams, \ie child nodes, are
referred to as \emph{compound nodes} (as opposed to \emph{atomic
nodes}); a graph that contains compound nodes is a \emph{compound
graph}. We refer to the ports of a compound node as
\emph{hierarchical ports}. These can be used to connect atomic nodes
inside a compound node to atomic nodes on the outside.

The main additional requirements for layout of data flow 
diagrams on top of standard graph drawing conventions are 
therefore~\cite{SpoenemannFvH+09}:
(R1) clearly visible flow,
(R2) ports and port constraints, 
(R3) compound nodes, 
(R4) hierarchical ports, 
(R5) orthogonal edge routing, and
(R6) orthogonalized node positions to emphasize R1 using horizontal edges.

The starting point for our approach is constrained stress 
majorization~\cite{DwyerKM06b}. This extends the original
stress majorization model~\cite{GansnerKN05} to support separation constraints
that can be used to declaratively enforce 
node alignment, non-overlap of nodes, flow in directed graphs,
and to cluster nodes inside non-overlapping regions.
Brandes \etal~\cite{BrandesEKW02} provide one method to orthogonalise
an existing layout based on the topology-shape-metrics approach, 
but in order to handle requirements R1--6 we instead use the
heuristic approach of Kieffer \etal~\cite{KiefferDM+13} to apply alignment
constraints within the stress-based model.

Our \acf{codaflow} layout algorithm is a pipeline with three stages:
\begin{enumerate}
  \item Constrained Stress-Minimizing Node Positioning
  \item Grid-Like Node Alignment
  \item Orthogonal Edge Routing
\end{enumerate}
The intermediate results of this pipeline are depicted in
\autoref{fig:pipeline}.
Single stages can be omitted, \eg when no edge routing is
required or initial node positions are given.
In this section we restrict our attention to flat graphs, \ie those without
compound nodes, while
\autoref{sec:compound_graphs}~extends the ideas to compound graphs.

\begin{figure}[tbp]
  \centering
  \subfloat[]{
    \includegraphics[width=.3\textwidth]{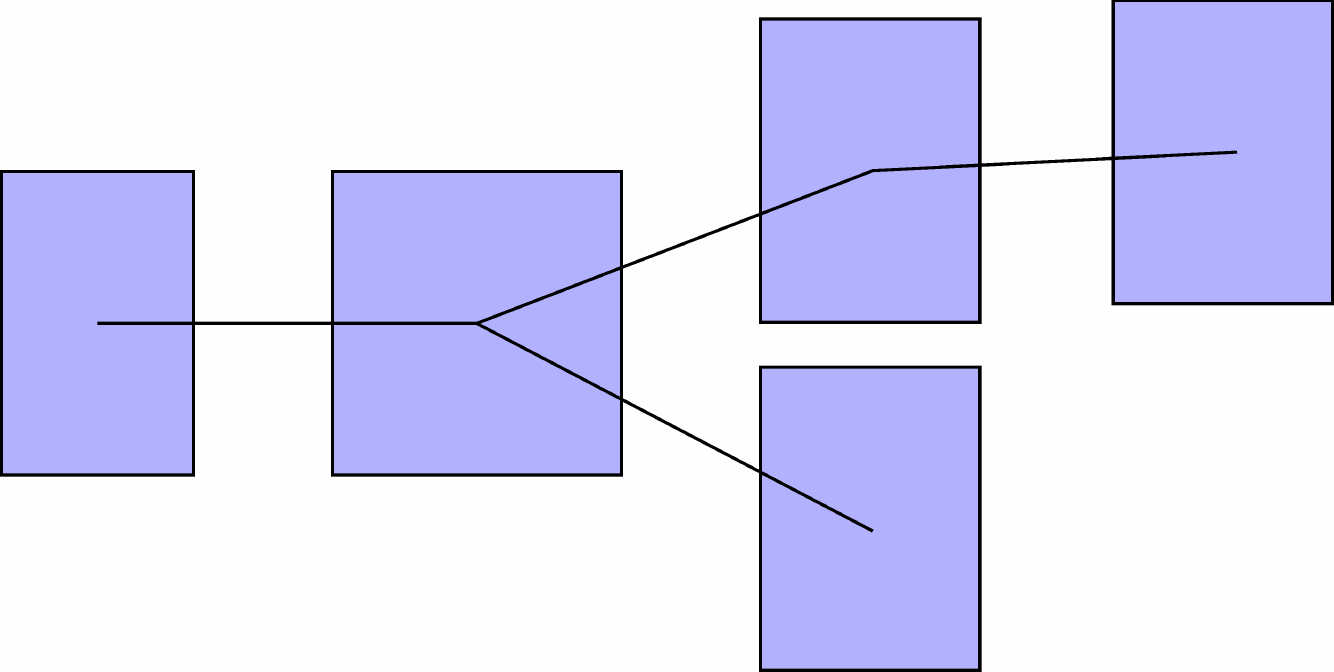} 
  }\hspace*{3em}
  \subfloat[]{
    \includegraphics[width=.40\textwidth]{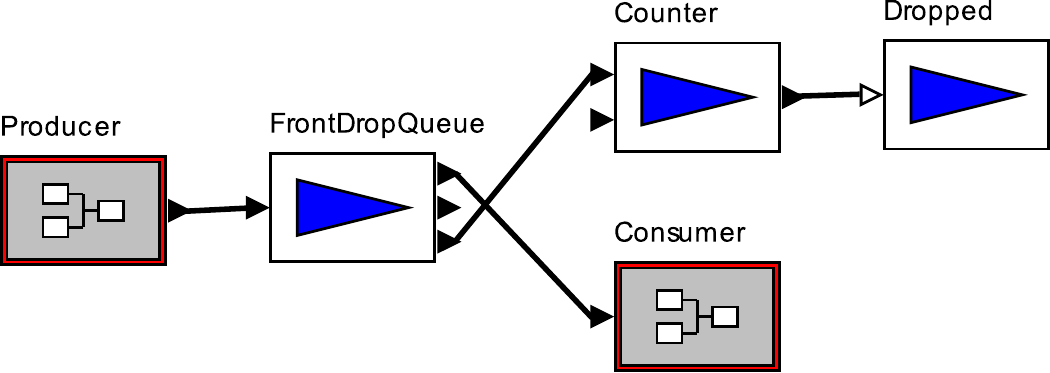} 
  }\\
  \subfloat[]{
   \label{fig:port_dummies_internal_rep}
   \includegraphics[width=.3\textwidth]{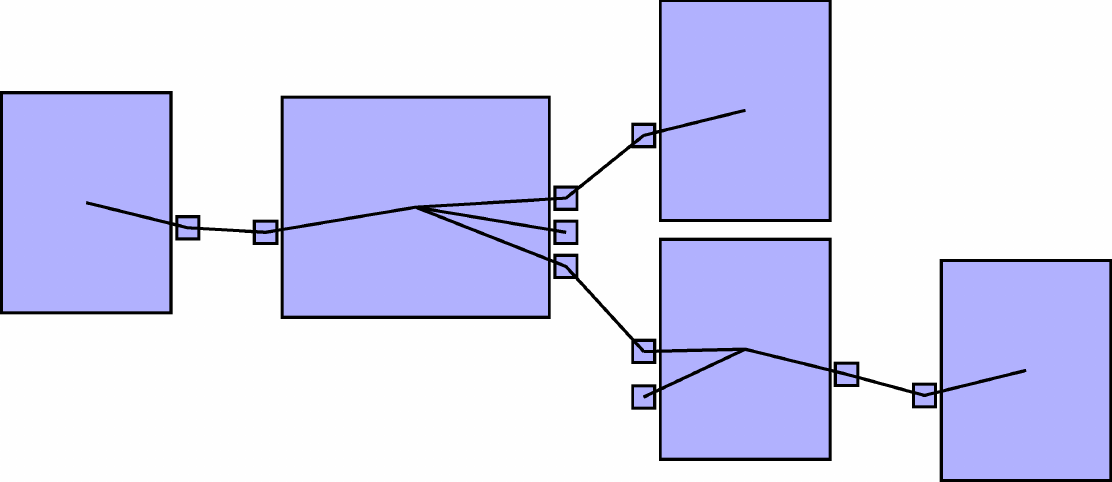} 
  }\hspace*{3em}
  \subfloat[]{
    \includegraphics[width=.40\textwidth]{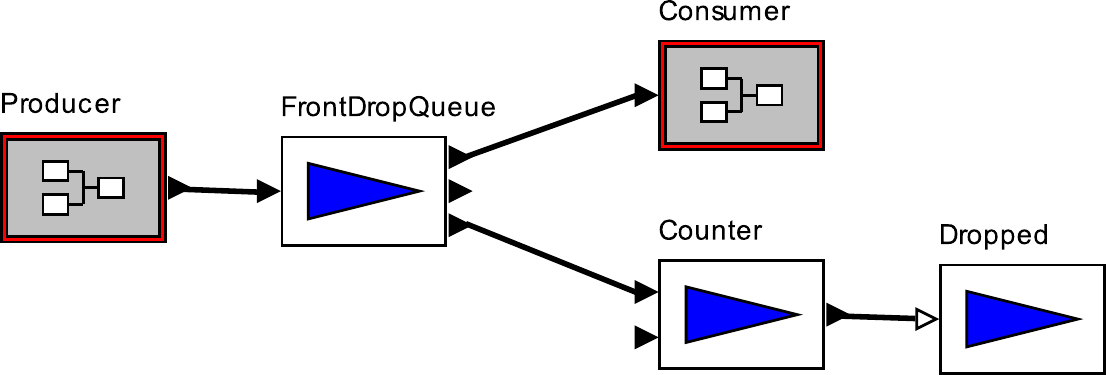} 
  }
  \caption{
    Awareness of ports is important to achieve good node positioning. 
    (a) and (c) show internal representations of what is passed to 
    the layout algorithm, (b) and (c) show the resulting drawings.
    (a) is unaware of ports and yields node positions that introduce
    an edge crossing in (b). In (c) ports are considered and the 
    unnecessary crossing is avoided in (d).
    Note, however, while the chance is higher that (c) is cross free, it
    is not guaranteed.
  }
  \label{fig:port_dummies}
\end{figure}

\subsection{Constrained Stress-Minimizing Node Positioning}
Traditional stress models for graph layout expect a simple graph %
without ports, so a key idea in order to handle data flow diagrams is to
create a small node to represent each port, called a
\emph{port node} or \emph{port dummy}, as in Fig.~\ref{fig:port_dummies_internal_rep}.
If $D$ is the set of all these, and $\delta:P\rightarrow D$ maps
each port to the dummy node that represents it, we construct a new graph
$G' = (V', E')$ where $V' = V \cup D$, and
\[
  E' = \{ ( \delta(p_1), \delta(p_2) ) : (p_1,p_2) \in E \} \cup
       \{ ( \pi(p), \delta(p) ) : p \in P \}
\]
includes one edge representing each edge of the original graph, and an edge
connecting each port dummy to its parent node.
We refer to the $v \in V$ as \emph{proper} nodes.

Depending on the specified port constraints (R2) we 
restrict the position of each port dummy $\delta(p)$ relative to its
parent node $\pi(p)$ using separation constraints.
For instance, for a rigid relative position
we use one separation constraint in each dimension,
whereas we retain only the $x$-constraint if $\delta(p)$ need only
appear on the left or right side of $\pi(p)$.
The use of port nodes allows the constrained stress-minimizing
layout algorithm to untangle the graph while being aware of 
relative port positions, resulting in fewer crossings, as illustrated in
\autoref{fig:port_dummies}.

Our constrained stress-based
layout uses the methods of Dwyer \etal~\cite{DwyerKM06b} to minimize
the \emph{P-stress} function~\cite{DwyerMW09a}, a variant
of \emph{stress}~\cite{GansnerKN05} that does not penalise
unconnected nodes being more than their desired distance apart:
\begin{equation}\label{eq:pstress}
  \sum_{u < v \in V'} w_{u v}
    \left( \left( \ell p_{u v} - b(u,v) \right)^{+} \right)^{2} +
  \sum_{(u,v) \in E'} \ell^{-2}
    \left( \left( b(u,v) - \ell \right)^{+} \right)^{2}
\end{equation}
where $b(u,v)$ is the Euclidean distance between the boundaries of nodes $u$
and $v$ along the straight line connecting their centres,
$p_{u v}$ the number of edges on the shortest path between nodes $u$ and $v$,
$\ell$ an ideal edge length,
$w_{u v} = (\ell p_{u v})^{-2}$, and
$(z)^+=\max(z,0)$.

\paragraph{Ideal Edge Lengths.}
Instead of using a single ideal edge
length $\ell$ as in \eqref{eq:pstress},
which can result in 
cluttered areas where multiple nodes are highly connected,
we may assign custom edge lengths $\ell_{u v}$,
choosing larger values to separate such nodes. 
In \autoref{fig:port_dummies} the ideal edge
lengths of the two outgoing edges of the 
\texttt{FrontDropQueue} actor 
are chosen slightly larger than for 
the two other edges.

The length of the edge $(\pi(p),\delta(p))$
connecting a port dummy to its parent node
is set to the exact distance from the 
node's center to the port's center.

\paragraph{Emphasizing Flow.}
A common requirement for data flow diagrams is that the majority of
edges point in the same direction (here left-to-right).
For this we introduce separation constraints for 
edges ($u$, $v$) of the form 
$x_u + g \leq x_v$, where $g > 0$ is a pre-defined spacing value,
ensuring that $u$ is placed left of $v$.
We refer to these constraints as \emph{flow constraints}.

Special care has to be taken for cycles, as they would introduce
contradicting constraints. We experimented with different 
strategies to handle this. 1) We introduced the constraints
even though they were contradicting (and let the solver choose which
one(s) to reject); 2) We did not generate \emph{any} flow
constraints for edges that are part of a strongly connected
component; 3) We employed a
greedy heuristic by Eades~\etal~\cite{EadesLS93} (known 
from the layer-based approach)
to find the minimal feedback arc set, and withheld flow
constraints for the edges in this set. 
Our experiments showed that the third strategy yields the best results.

\paragraph{Execution.}
We perform three consecutive layout runs, iteratively adding
constraints: 1) Only port constraints are applied, allowing
the graph to untangle and expose symmetry; 2) Flow 
constraints are added, but overlaps are still allowed so that
nodes can float past each other, swapping positions where 
necessary; 3) Non-overlap constraints are applied to separate all nodes as desired.

\subsection{Grid-like Node Alignment}

While yielding a good distribution of nodes overall, stress-minimization tends to produce
an organic layout with paths splayed at all angles, which is
inappropriate for data flow diagrams. The layout needs to be
\emph{orthogonalized}, \ie~connected nodes brought into alignment with one
another so that where possible edges form straight horizontal lines, 
visually emphasizing horizontal flow.

\begin{SCfigure}[30][tbp]
  \centering
    \vspace{3em}
  \subfloat[Proper nodes connected via port nodes]{
    \includegraphics[width=.2\textwidth]{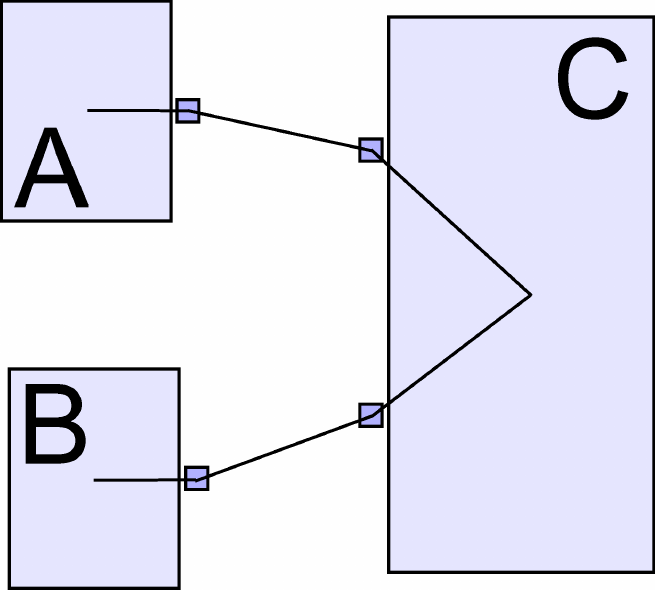} 
    \label{fig:dummy_nodes}
  }\hspace*{2em}
  \subfloat[Ports aligned by ACA]{
   \includegraphics[width=.2\textwidth]{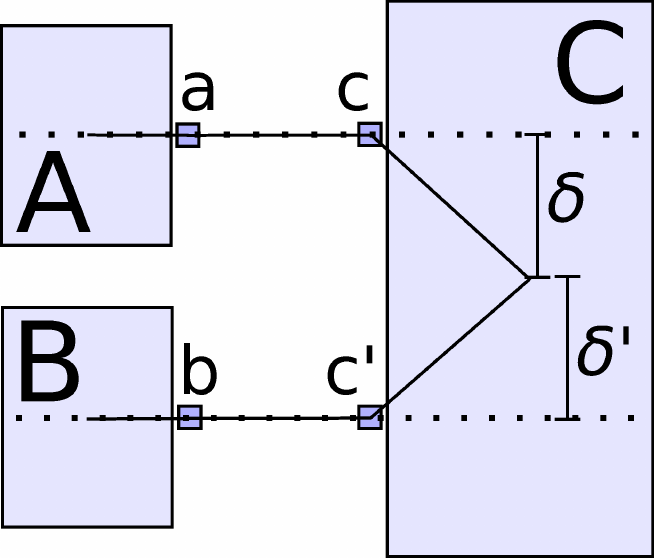} 
   \label{fig:aca_ports_aligned}
  }
  \caption{
  {\small
  In the new port model, two proper nodes may be connected to the same side
  of another via ports, as in (a).
  The systematic use of \emph{offset alignments} between port nodes and their parents,
  \ie~constraints of the form $y_{\delta(p)} + \delta = y_{\pi(p)}$, $\delta \neq 0$
  as shown in (b), creates a risk of node-edge and node-node overlaps far exceeding what was
  anticipated with the original \acs{aca} algorithm, as could have occurred
  here had node $B$ been as tall as node $C$, for example.
  We have extended \acs{aca} to properly handle such cases.
  }
  }
  \label{fig:aca}
\end{SCfigure}

For this purpose we apply the \acf{aca} algorithm~\cite{KiefferDM+13}. 
Since it respects existing flow constraints, it only attempts to align
edges horizontally.
However, our replacement of the given graph $G$
by the auxiliary graph $G'$ with port nodes tends to subvert
the original intentions of \acs{aca}, so it requires some adaptation.
Whereas the original \acs{aca} algorithm expected at most one proper
node to be aligned with another in a given compass direction,
in our case (with ports)
it will often be desirable to have more.
See Fig.~\ref{fig:aca}.

In order to adapt \acs{aca} to the new port model we made it possible to
ignore certain edges---namely those connecting port nodes to their
parents---and also generalised its overlap prevention methods significantly.
Instead of the simple procedure for preventing multiple alignments in a single
compass direction~\cite{KiefferDM+13}, we use the VPSC solver~\cite{DwyerMS06}
for trial satisfaction
of existing constraints, the new potential
alignment, as
well as non-overlap constraints between all nodes and a dummy node representing the
potentially aligned edge.

Thus, while the \acs{aca} process continues to merely centre-align nodes---in
this case port nodes $d \in D$---we have allowed it to \emph{in effect} align several proper
nodes $v_1, \ldots, v_k \in V$ with a single one $u \in V$ at port
positions as in~\autoref{fig:aca}, meeting the requirement R6 of data flow diagrams.

\subsection{Edge Routing}
We now consider node positions
to be fixed, and use the methods of Wybrow \etal~\cite{WybrowMS10} 
to route the edges orthogonally.
We return from $G'$ to $G$, using the final positions of the port
nodes $d \in D$ to set \emph{routing pins}, fixed port positions
on the nodes $v \in V$ where the edges should
connect.

\section{Handling Compound Graphs}
\label{sec:compound_graphs}
\begin{figure}[tb]
  \centering
  \subfloat[Layout with layer-based methods by Schulze~\etal]{
    \includegraphics[width=.7\textwidth]{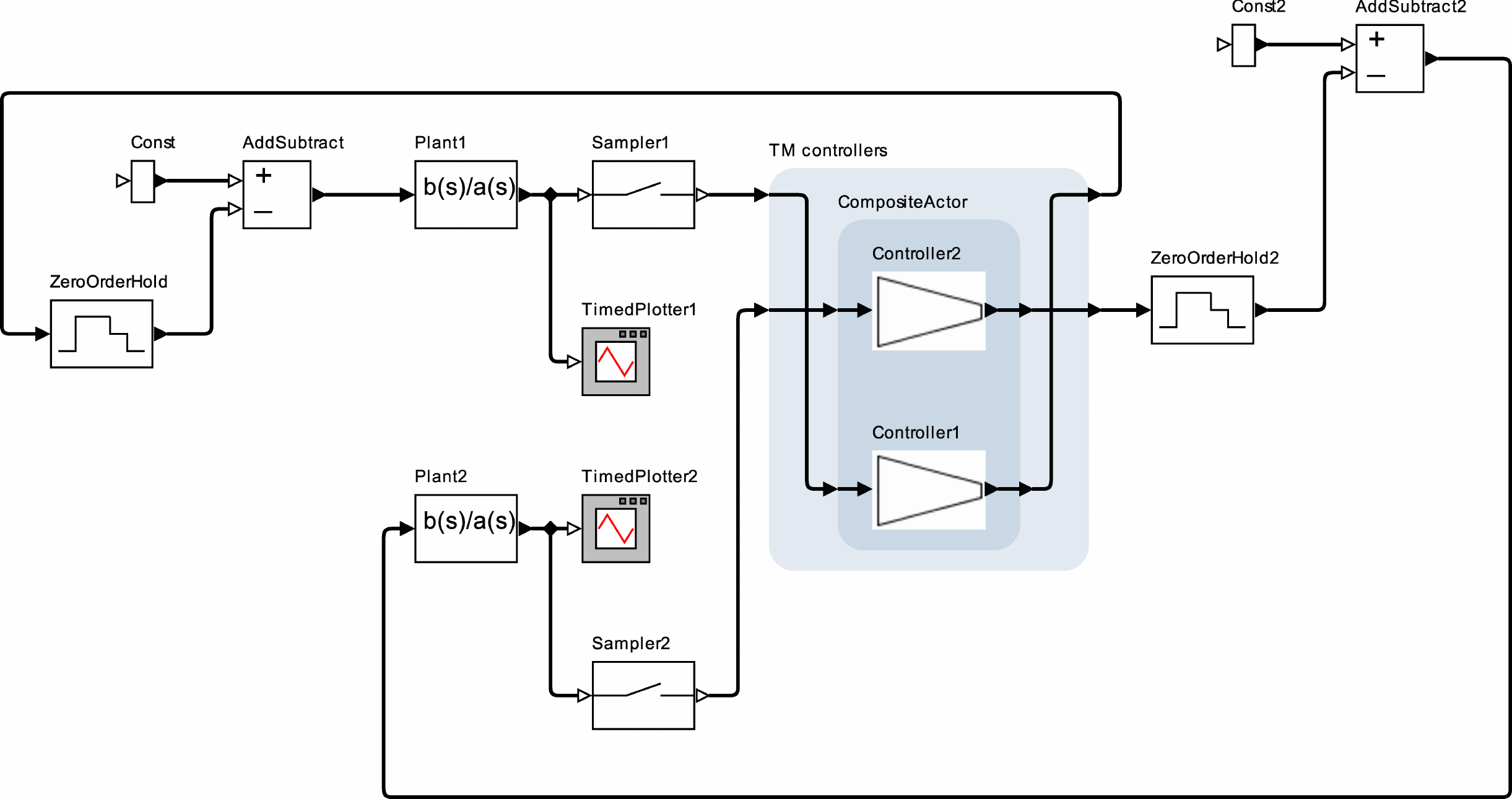} 
    \label{fig:unconnected_klay}
  }\\
  \subfloat[Layout with the CoDaFlow algorithm presented here.]{
    \includegraphics[width=.6\textwidth]{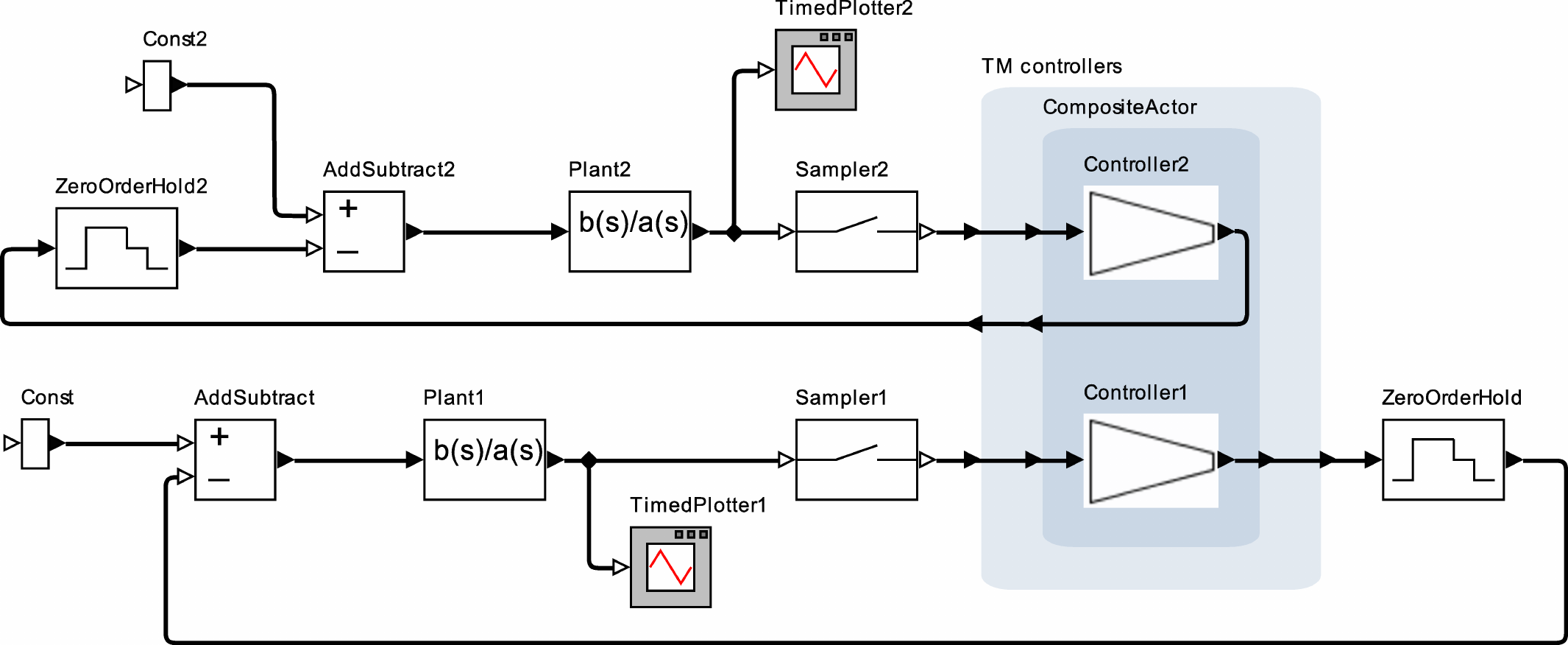} 
    \label{fig:unconnected_cola}
  }
  \caption{
    Two layouts of the same Ptolemy diagram.
    While two distinct networks are interleaved in (a),
    they are clearly separated and the two crossings are avoided in (b).
  }
  \label{fig:unconnected}
\end{figure}

When handling compound graphs, different strategies for dealing with
compound nodes. Schulze~\etal employ a 
\emph{bottom-up} strategy, treating every compound node as
a separate graph, starting with the inner-most nodes.
This allows application of different layout algorithms to each
subgraph which reduces the size of the layout problem,
and possibly the overall execution time.
They remark, however, that the procedure can yield unsatisfying 
layouts since the surroundings of a compound node are not known;
see \autoref{fig:unconnected_klay} for an example where two unnecessary 
crossings are created inside the \texttt{TM controllers} actor
and two separate networks are interleaved. 
A \emph{global} approach would solve this issue, positioning
all compound nodes along with their children at the same time.

Even though we focus our attention on a global 
approach in what follows, our methods are flexible in that we 
may choose between a bottom-up and a global 
strategy in each stage of our pipeline.

A compound graph $G$ is transformed into $G'$ as above, which is
used to construct a \emph{flat} graph 
$G'' = (A, E'')$ where $A \subseteq V'$ is the set of atomic nodes
and their port nodes, and $E'' = U \cup H$ with
\begin{align*}
  U = & \{ (\delta(p_1), \delta(p_n)) : \pi(p_1), \pi(p_n) \in A  \} \\
  H = & \{ ( \delta(p_1), \delta(p_n) ) : 
    \exists (p_1, p_2),(p_2,p_3),\dots,(p_{n-2}, p_{n-1}),(p_{n-1}, p_n) \in E : \\
      &       \pi(p_1), \pi(p_n) \in A \wedge \pi(p_2),\dots,\pi(p_{n-1}) \in V \backslash A \}
\end{align*}

Intuitively, compound nodes are neglected along with their ports
and only atomic nodes are retained.
Sequences of edges that span hierarchy boundaries, 
\eg the three edges between \texttt{Sampler2} 
and \texttt{Controller2} in \autoref{fig:unconnected_cola}, 
are replaced by a single edge that directly connects the two 
atomic nodes. Note that for hyperedges 
multiple edges have to be created.
\emph{Cluster constraints} guarantee that children of  
compound nodes are kept close 
together and are not interleaved with any other nodes.
For instance, the \texttt{CompositeActor}
in \autoref{fig:unconnected_cola} yields 
a cluster containing \texttt{Controller1} and \texttt{Controller2}.

To return to $G$, the clusters' dimensions,
\ie their rectangular bounding boxes, are 
applied to the compound nodes in $V\backslash A$.
The edges in $H$ are split 
into segments $s_1,\dots,s_n$ based on the crossing points 
$c_i$ with clusters.
The route of $s_i$ is applied to the corresponding 
edge $e \in E$
and the $c_i$ determine the positions of the hierarchical ports.

\section{Evaluation and Discussion}
\label{sec:discussion}

We evaluate our approach on a set of data flow diagrams
that ship with the Ptolemy 
project\footnote{\url{http://ptolemy.eecs.berkeley.edu/}},
comparing with the \acs{klay} layered algorithm of
Schulze~\etal
Diagrams were chosen to be roughly the size Klauske 
found to be typical for real-world Simulink models
from the automotive industry~\cite{Klauske12} 
(about 20 nodes and 30 edges per hierarchy level).

\paragraph{Metrics.}

Well established metrics to assess the quality of a drawing 
are \emph{edge crossings} and \emph{edge bends}~\cite{Purchase97},
two metrics directly optimized by the layer-based approach.
More recently, \emph{stress} and \emph{edge length variance} 
were found to have a significant impact on the readability of a 
drawing~\cite{DwyerLFQ+09}.
Additionally, we regard compactness in terms of 
\emph{aspect ratio} and \emph{area}.

So that comparisons of edge length and of layout area can be meaningful,
we set the same value for \acs{klaylay}'s inter-layer distance and
\acs{codaflow}'s ideal separation between nodes.

The $P$-stress of a given (already layouted) diagram depends on the choice of the ideal edge
length $\ell$ in \eqref{eq:pstress}, and the canonical choice
$\bar\ell$ is that where the function takes its global minimum.
If $L$ is a list of all the individual ideal lengths $\ell_{uv} = b(u,v)/p_{uv}$, then
$\bar\ell$ is equal to the \emph{contraharmonic mean} $C(L_j)$
(\ie~the weighted arithmetic mean in which the weights equal the values)
over a certain sublist $L_j \subseteq L$.
Namely, if $L_E = \langle \ell_{uv} : (u,v) \in E \rangle$
and $L \setminus L_E = \langle \ell_1 \leq \ell_2 \leq \cdots \leq \ell_\nu \rangle$,
then $L_j = L_E \cup \langle \ell_1, \ell_2, \ldots, \ell_j \rangle$ for some
$0 \leq j \leq \nu$.
Since $\nu$ is finite, we can compute each $C(L_j)$ and take $\bar\ell$ to be
that at which the $P$-stress is minimized. See Appendix~\ref{sec:pstress}.
\newcommand{\csep}{c@{\hskip 1em}}
\newcommand{\mamhead}{ \enskip Min \quad Avg \quad Max }
\newcommand{\mam}[3]{ \enskip #1 \quad #2 \quad #3 }

\begin{table}[tb]
  \scriptsize
  \centering
  \caption{
   Evaluations of 110 flat diagrams with 10--23 nodes (9--30 edges) and
   10 compound diagrams with 12--38 nodes (12--52 edges).
   Figures for stress, average edge length, variance in edge length,
   and area are given as the ratio of \acs{codaflow} divided by \acs{klay}. 
   Values below 1 indicate a better performance of \acs{codaflow}.
   An average value shows the general tendency while minimal and maximal 
   values show the best and worst performance.
  }
  \begin{tabular}{l \csep | \csep | \csep | c}
    \toprule
                 & \enskip Stress & \enskip EL Variance & \enskip EL Average & \enskip Area \\
                 & \mamhead & \mamhead & \mamhead & \mamhead \\ 
    \midrule
      Comp.         & \mam{0.27}{0.75}{0.97} %
                    & \mam{0.11}{0.29}{0.61} %
                    & \mam{0.39}{0.57}{0.79} %
                    & \mam{0.50}{0.88}{1.28} \\ %
      \midrule
      Flat          & \mam{0.34}{0.77}{1.13} %
                    & \mam{0.03}{0.60}{1.92} %
                    & \mam{0.34}{0.84}{1.10} %
                    & \mam{0.62}{1.11}{2.01} \\ %
    \bottomrule
  \end{tabular}
  \label{tab:eval}
\end{table}

\begin{table}[tb]
  \scriptsize
  \centering
  \caption{
   Results for the metrics aspect ratio, crossings, and average
   bends per edge. As opposed to \autoref{tab:eval}, figures 
   are absolute values.
  }
  \begin{tabular}{l l | \csep | \csep | c}
    \toprule
                          &   &  \enskip Aspect Ratio & \enskip Crossings & \enskip Bends/Edge \\ 
                          &   &  \mamhead     & \mamhead  & \mamhead \\
    \midrule
    \multirow{2}{*}{Comp.}  & {\tiny \acs{codaflow} \enskip}  
                               & \mam{1.27}{1.83}{2.51} %
                               & \mam{0.00}{3.40}{10.0} %
                               & \mam{0.92}{1.25}{1.56} \\ %
                            & {\tiny \acs{klay}}     
                               & \mam{1.51}{2.76}{4.94} %
                               & \mam{0.00}{1.20}{6.00} %
                               & \mam{0.68}{0.97}{1.22} \\ %
    \midrule
    \multirow{2}{*}{Flat}   & {\tiny \acs{codaflow} \enskip}  
                               & \mam{0.32}{2.47}{5.96} %
                               & \mam{0.00}{1.25}{11.0} %
                               & \mam{0.42}{1.16}{2.31} \\ %
                            & {\tiny \acs{klay}}     
                               & \mam{0.37}{2.77}{9.00} %
                               & \mam{0.00}{1.02}{7.00} %
                               & \mam{0.22}{1.04}{1.73} \\ %
    \bottomrule
  \end{tabular}
  \label{tab:eval2}
\end{table}

\paragraph{Results.} 
\autoref{tab:eval} and~\ref{tab:eval2} show
detailed results for layouts created by \acs{codaflow}
and \acs{klaylay}.
We used two variations of the 
Ptolemy diagrams: small flat diagrams and compound diagrams 
(\cf examples in the appendix). 

For flat diagrams \acs{codaflow} shows a better performance on stress,
average edge length, and variance in edge length. 
\acs{codaflow} produced slightly more crossings, 
bends per edge and slightly increased area. 

More interesting are the results for the compound diagrams, which show
more significant improvements.  On average, \acs{codaflow}'s diagram
area was 88\% that of \acs{klaylay}, and edge length variance was only
29\%.  Also, the average aspect ratio shifts closer to that of
monitors and paper.  However, there is an increase in crossings.
Currently our approach does not consider crossings at all,
thus the increased average.
As can be seen in \autoref{fig:example}, the small number of additional
crossings are not ruinous to diagram readability,
and they could be easily avoided
by introducing further constraints, as discussed in
\autoref{sec:conclusions}.

\begin{figure}[tb]
  \centering
  \subfloat[
    Overall
  ]{
    \includegraphics[width=0.47\textwidth]{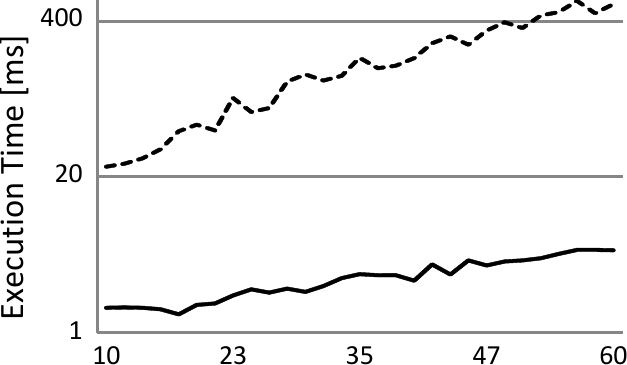}
  } \hspace*{0.7em}
  \subfloat[
    Pipeline Stages
 ]{
    \includegraphics[width=.47\textwidth]{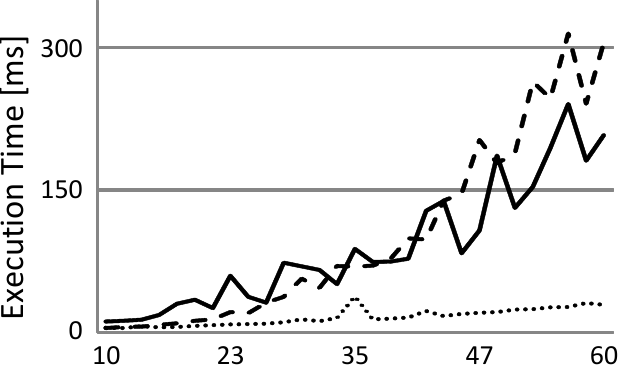}
  }
  \caption{
    Execution time plotted against the number of nodes $n$. For each $n$ 10 graphs were 
    generated randomly with an average of $1.5$ outgoing edges per node. 
    (a) Overall execution time of \acs{klaylay} (solid line) and \acs{codaflow} (dashed line).
    (b) Execution time of the pipeline steps: Untangling (solid line), 
    Alignment (dashed line), and Edge Routing (dotted line).
    Timings were conducted on an Intel i7 2.0 GHz with 8 GB RAM.
 }
  \label{fig:execution_time_chart}
\end{figure}

\paragraph{Execution Times.}
As seen in \autoref{fig:execution_time_chart}, the current \acs{codaflow}
implementation performs significantly slower than \acs{klaylay}, but it
still finishes in about half a second even on a large diagram of 60 nodes.
There is room for speedups,
for instance, by avoiding re-initialization 
of internal data structures between pipeline stages.
In addition, we plan to improve the incrementality of constraint
solving in the \acs{aca} stage, as well as performing faster satisfiability
checks wherever full projections are not required.

\paragraph{Implementation and Flexibility.}
Compared to \acs{klaylay} our approach is both 
easier to understand and implement,
and more flexible in its application.

In addition to the five main phases of \acs{klaylay},
about 10 to 20 \emph{intermediate processes} of 
low to medium complexity are used during each layout run.
Dependencies between these units have to be carefully managed 
and the phases have to be executed in strict order, \eg
the edge routing phase requires all previous 
phases.  

\acs{codaflow} optimizes only one goal function and 
addresses the requirements of data flow diagrams 
by successively adding constraints to the optimization process.
While we divide the algorithm into multiple stages,
each stage merely introduces the 
required constraints.
\acs{codaflow}'s stages can be used 
independently of each other, \eg
to improve existing layouts. 
Also, users can fine-tune generated drawings
using interactive layout~\cite{DwyerMW09b} methods.

\section{Conclusions}
\label{sec:conclusions}

We present a novel approach to layout of data flow diagrams
based on stress-minimization.
We show that it is superior to previous approaches with respect to
several diagram aesthetics. Also, it is more
flexible and easier to implement.\footnote{Author Ulf R\"uegg has worked on both \acs{klaylay} and \acs{codaflow}.}

The approach can easily be extended
to further diagram types with similar
drawing requirements, such as the \acf{sbgn}. 
To allow interactive browsing of larger diagram instances, however, 
execution time has 
to be reduced, \eg by removing overhead from both the
implementation and the pipeline steps.
Avoiding the crossing in \autoref{fig:port_dummies} is currently 
not guaranteed. We plan to detect such obvious cases 
via ordering constraints. 
In addition to \ac{aca}, the use of topological improvement strategies~\cite{DwyerMW09a} 
could help to reduce the number of edge bends further where
edges are almost straight.

\paragraph{\textbf{Acknowledgements.}}
Ulf R\"uegg was funded by a
doctoral scholarship 
(FIT{\hyp}weltweit) of the German Academic
Exchange Service.  Michael Wybrow was supported by the Australian Research Council (ARC) Discovery Project grant DP110101390.

\bibliographystyle{splncs03}
\bibliography{cau-rt,pub-rts}

\pagebreak
\appendix

\section{Appendix}

\subsection{Flat Graphs}

\begin{figure}[tbhp]
  \centering
  \subfloat[\acs{codaflow}]{
    \parbox{.39\linewidth}{
      \centering
      \includegraphics[scale=0.24]{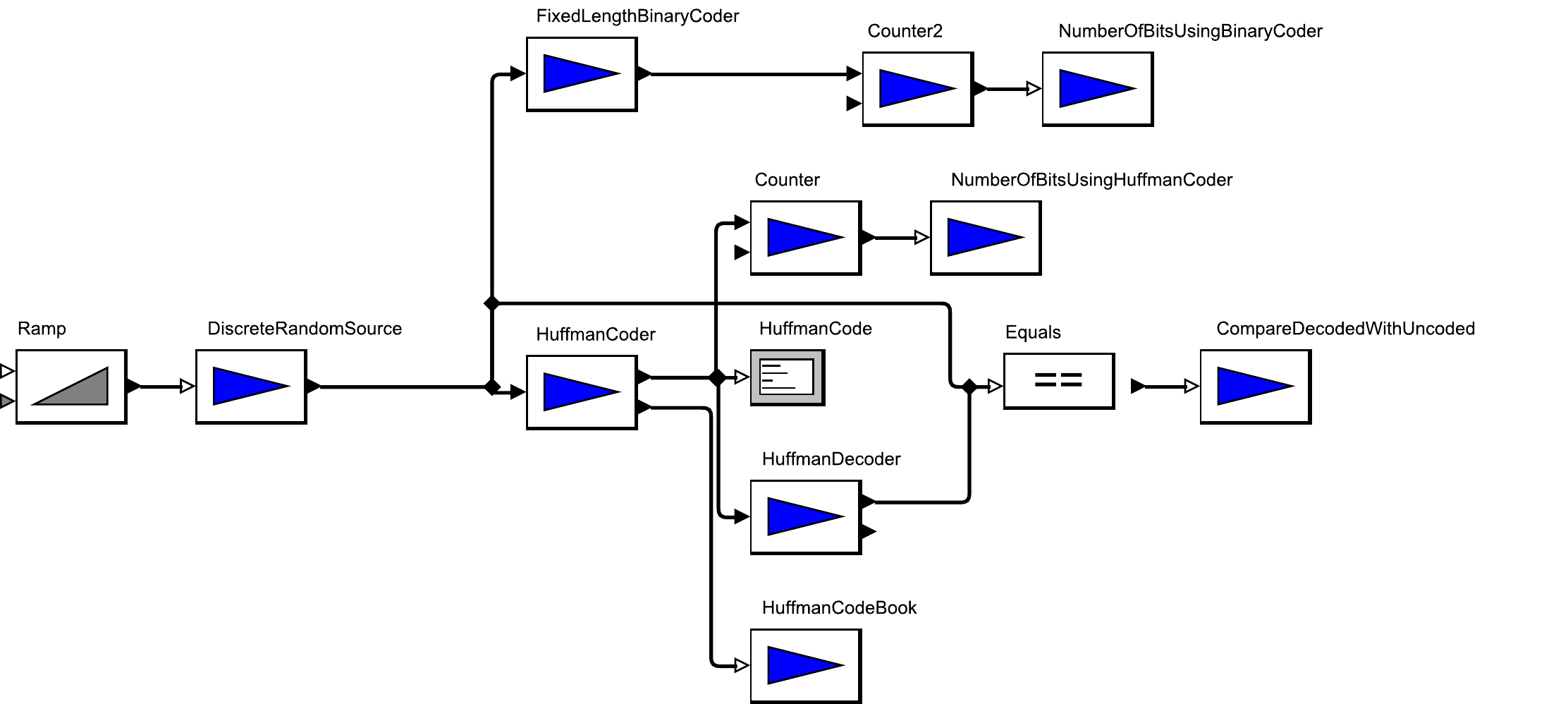} \\[3em]
      \includegraphics[scale=0.25]{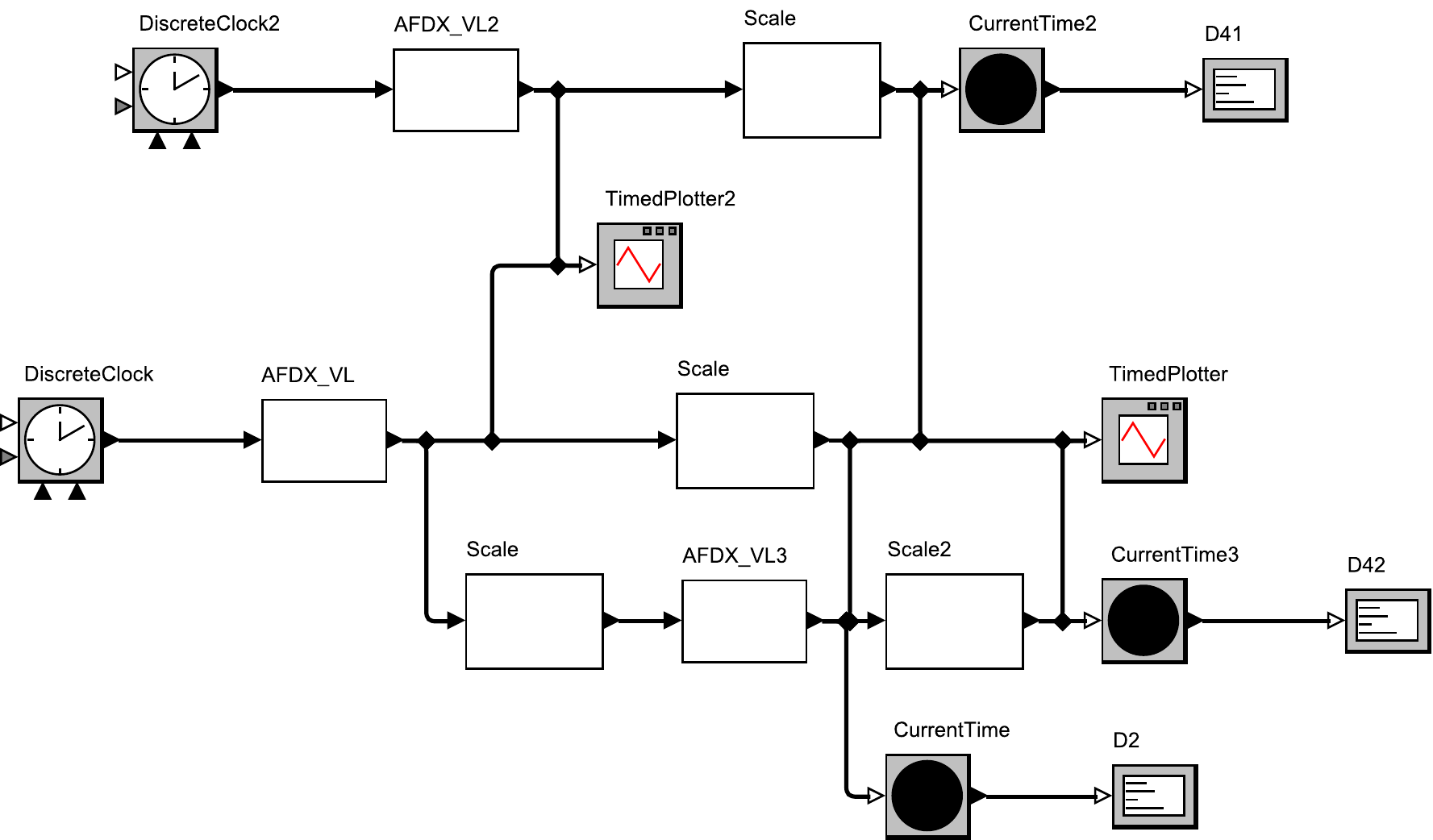} \\[3em]
      \includegraphics[scale=0.2]{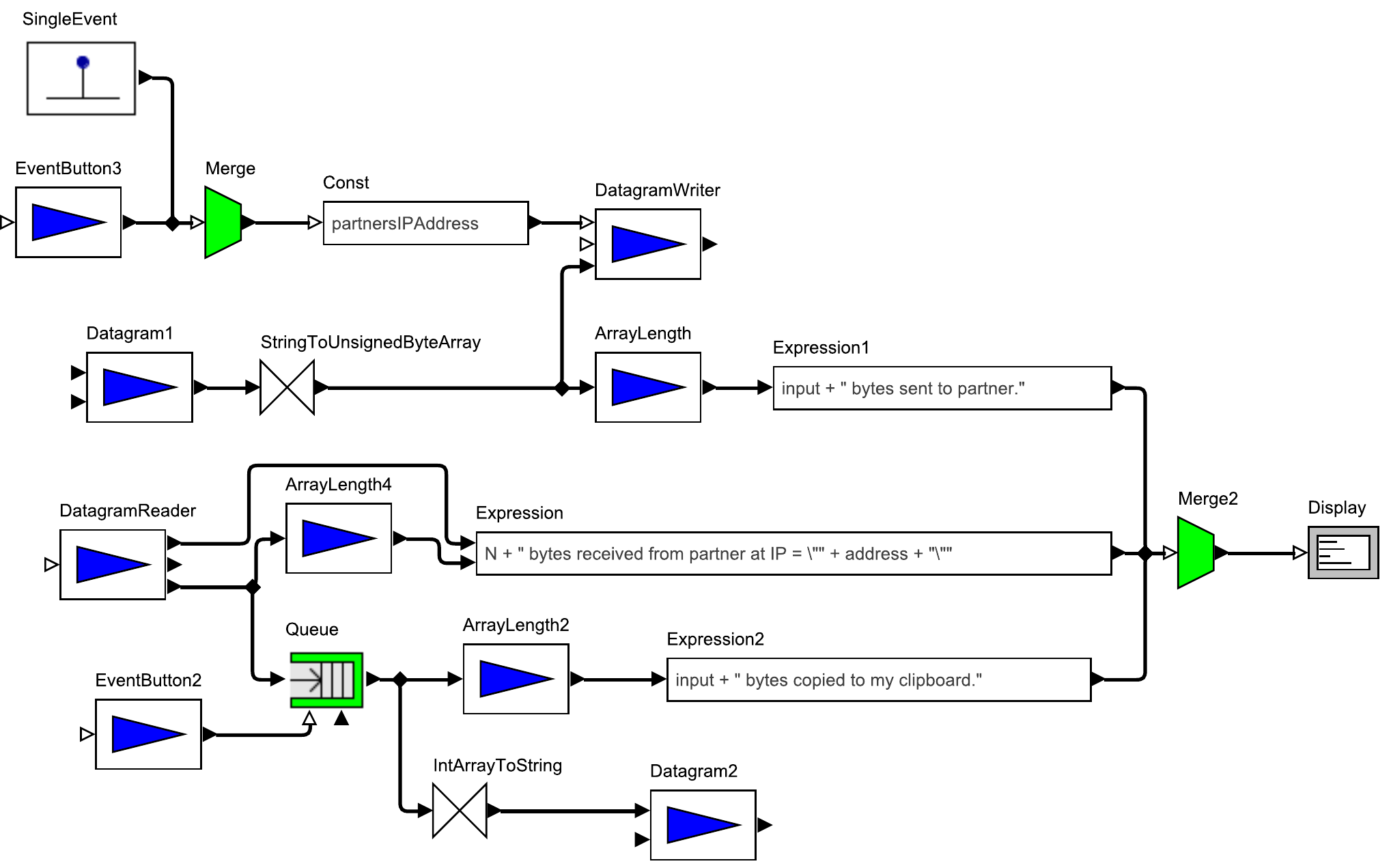}    
    }
  }
  \subfloat[\acs{klaylay}]{
    \parbox{.60\linewidth}{
      \centering
      \includegraphics[scale=0.24]{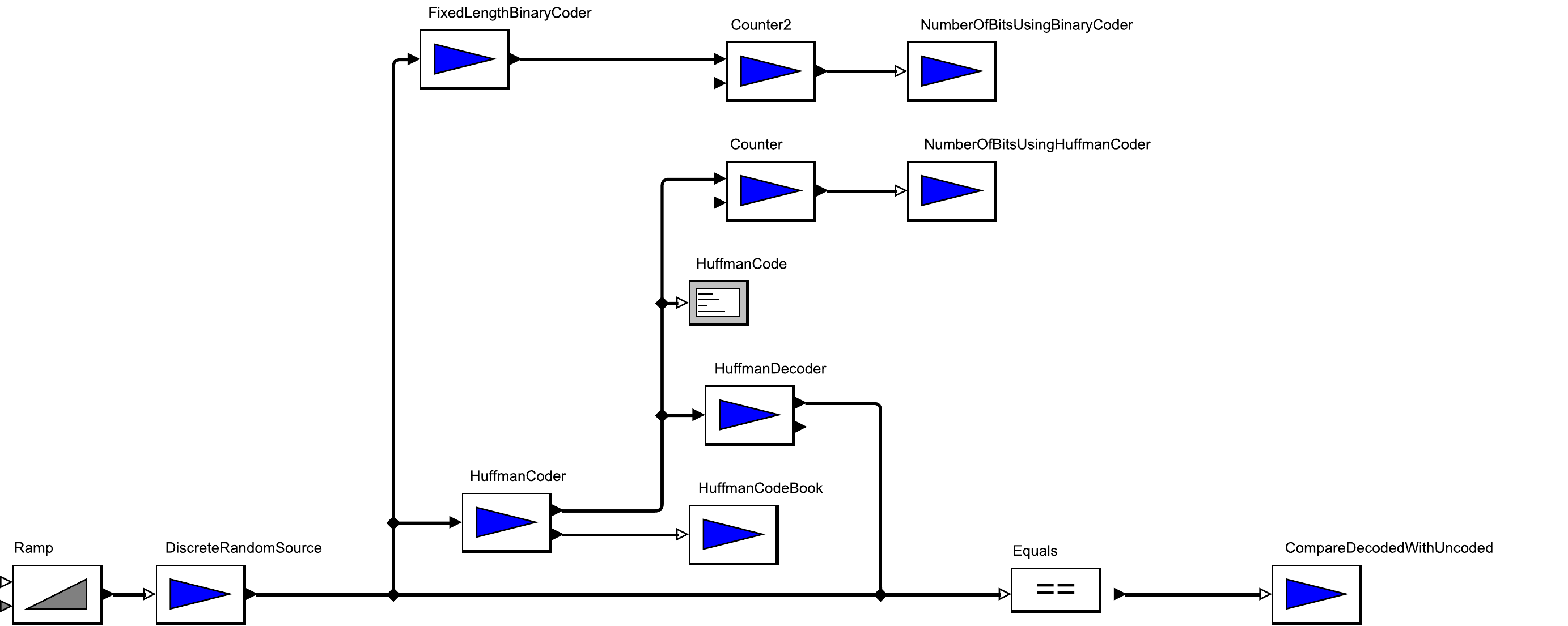} \\[3em] 
      \includegraphics[scale=0.25]{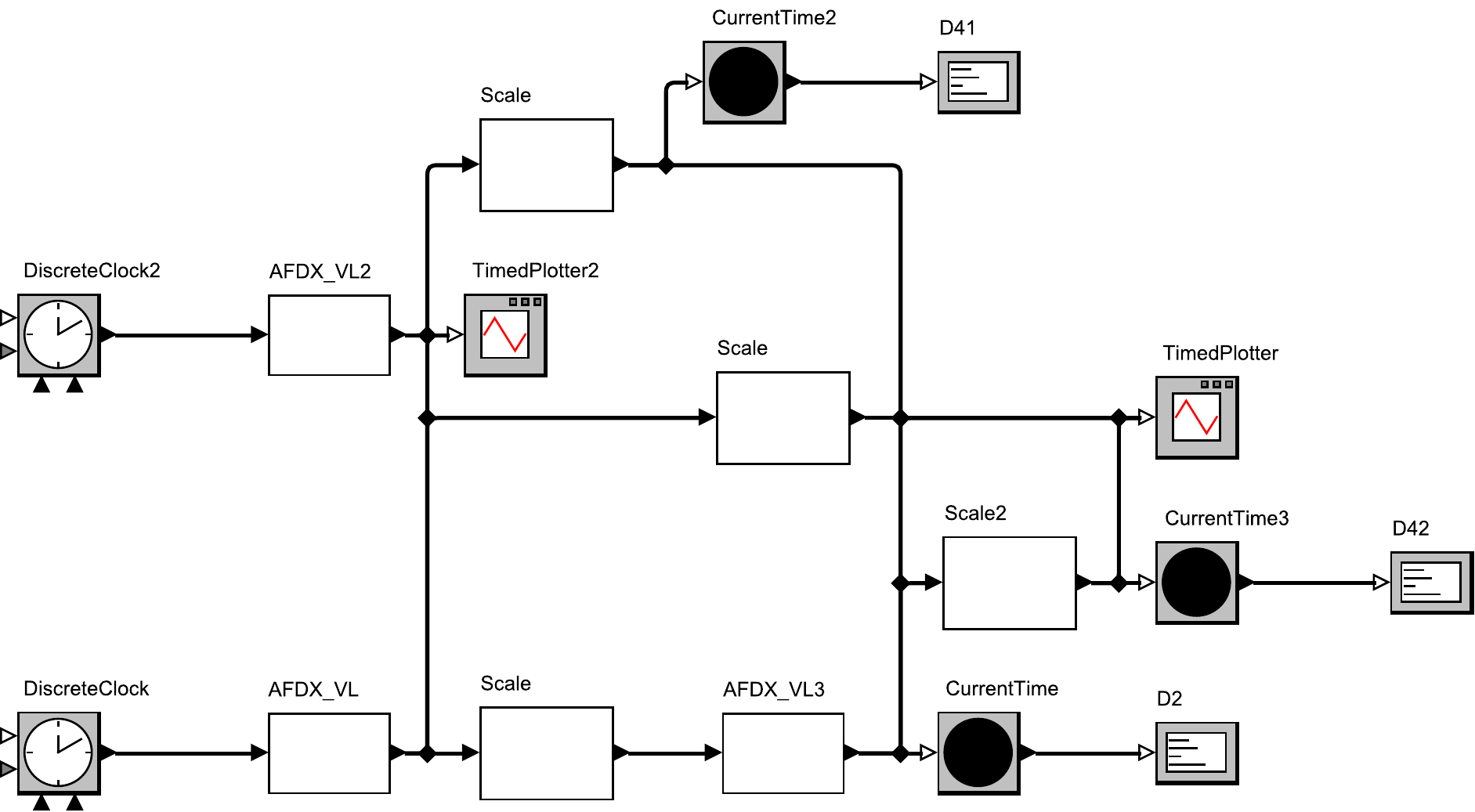} \\[3em]
      \includegraphics[scale=0.2]{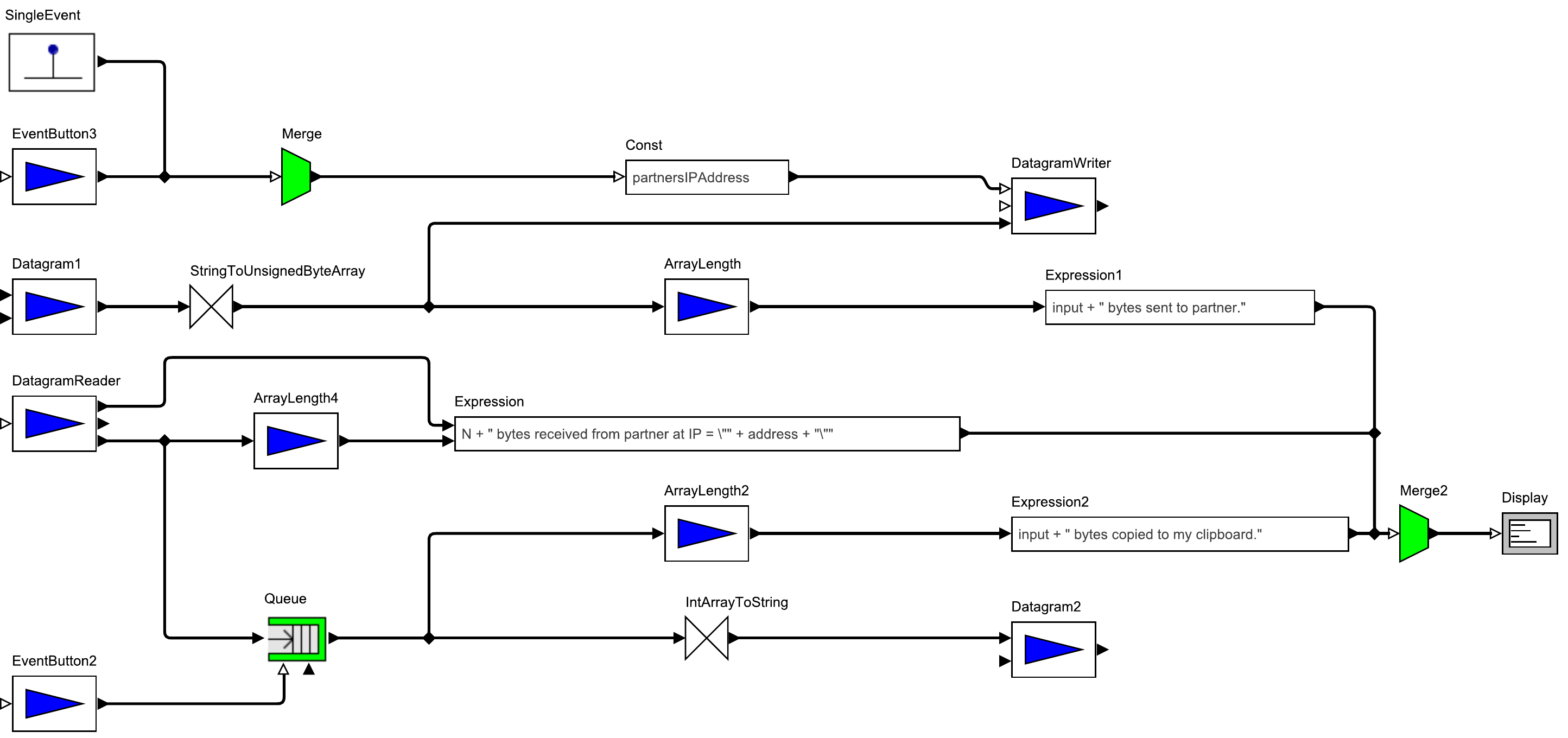} 
    }
  }
  \caption{
    Comparison of the generated layouts of \acs{codaflow} and \acs{klaylay} for flat graphs. 
    Diagrams are taken from the Ptolemy example library:
    \texttt{Huffman}, \texttt{AFDX}, and \texttt{Datagram} 
    (from top to bottom).
  }
\end{figure}

\newpage
\subsection{Compound Graphs}

\begin{figure}[tbh]
  \centering
  \subfloat[\acs{codaflow}]{
      \includegraphics[scale=0.38]{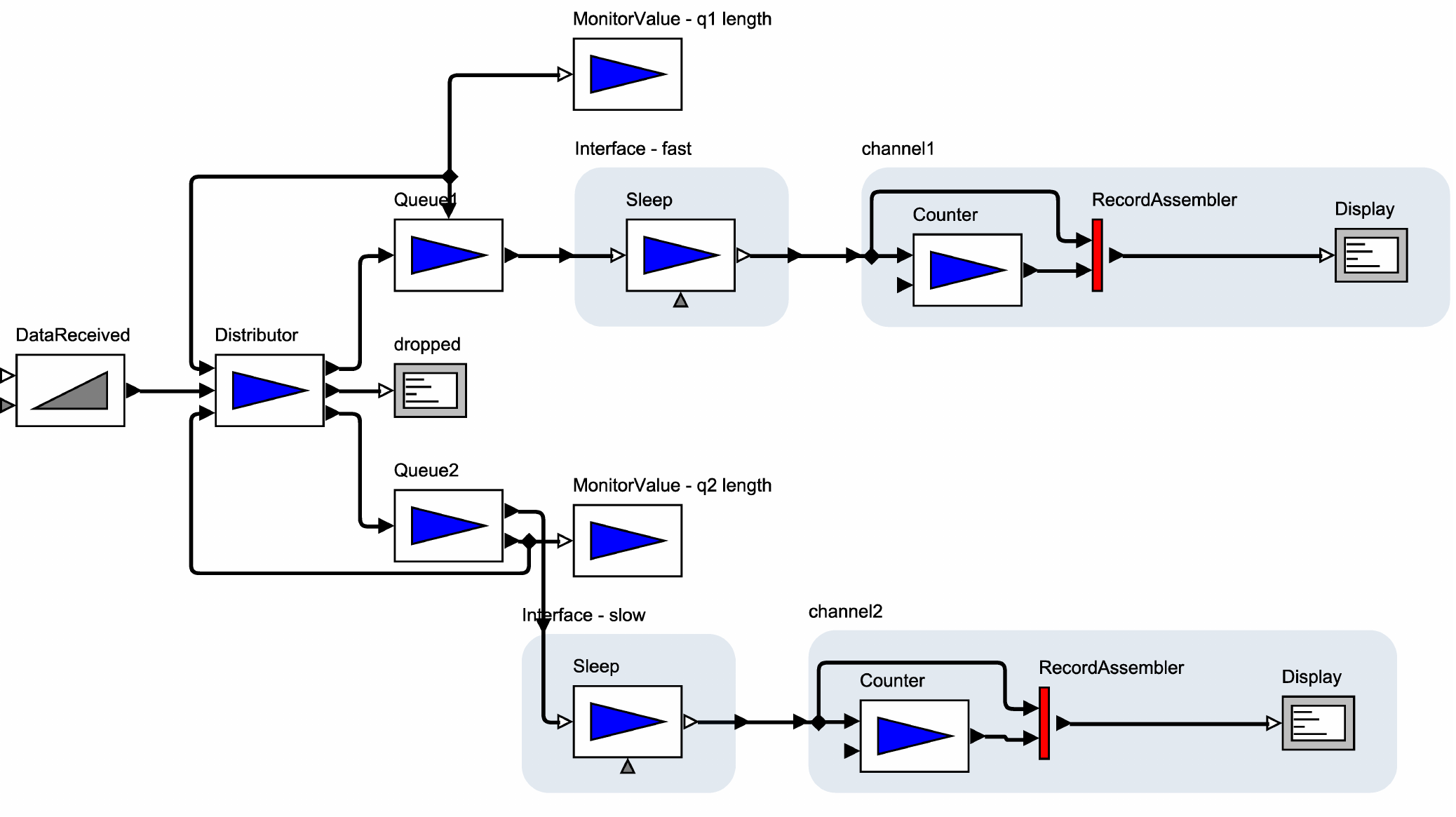}
  }\\
  \subfloat[\acs{klaylay}]{
      \includegraphics[scale=0.38]{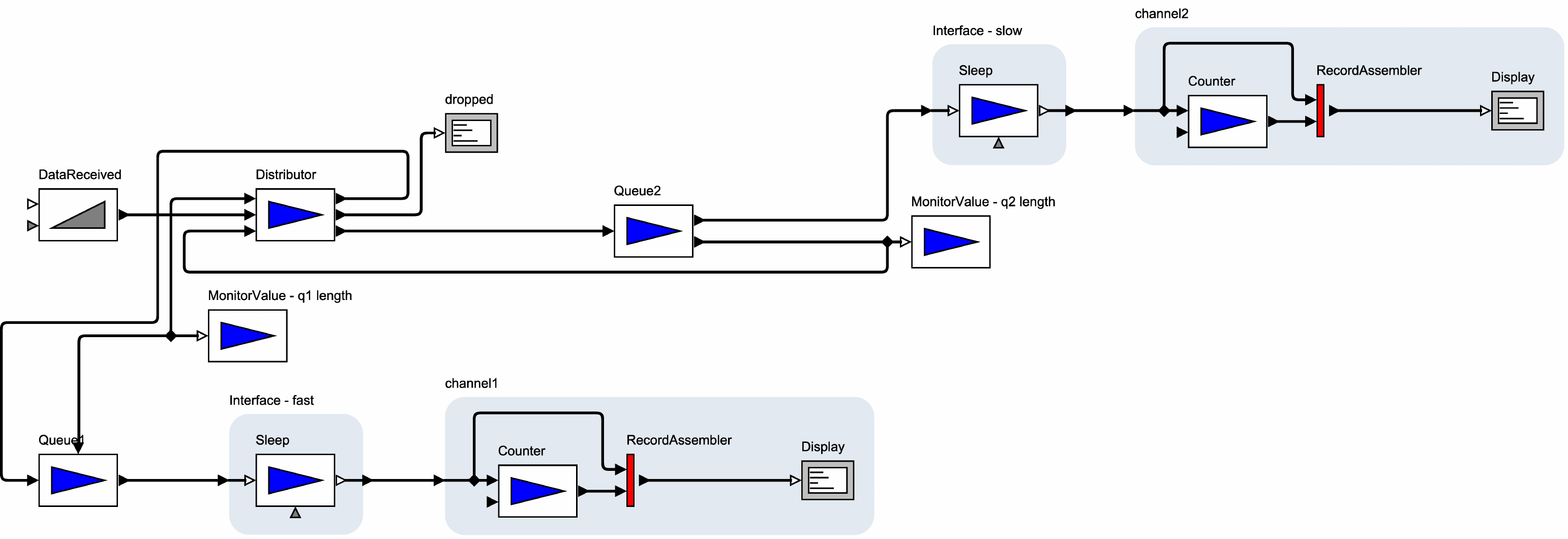}
  }
  \caption{
    Comparison of the generated layouts for 
    the \texttt{Router} diagram with 19 nodes, 24 edges, and 4 compound nodes.
  }
\end{figure}

\begin{figure}[t]
  \centering
  \subfloat[\acs{codaflow}]{
      \includegraphics[scale=0.35]{images/droprequest_cola}
  }\\
  \subfloat[\acs{klaylay}]{
      \includegraphics[width=0.98\textwidth]{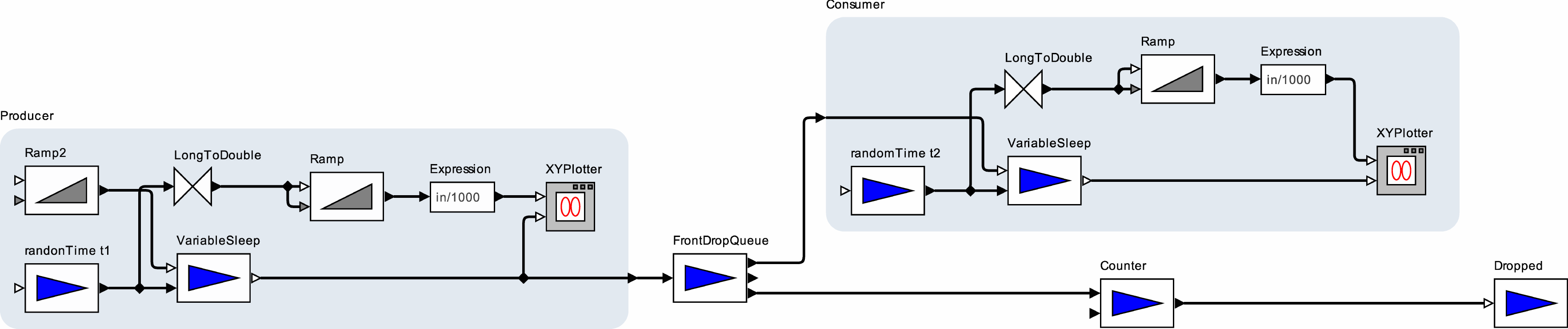}
  }
  \caption{
    Comparison of the generated layouts for
    the \texttt{dropqueuetest} diagram with 16 nodes, 21 edges, and 2 compound nodes. 
  }
\end{figure}

\newpage

\section{Computing ideal edge length to minimize $P$-stress}\label{sec:pstress}
%
%

Let a graph $G = (V,E)$ be given, along with some linear ordering on $V$.
We assume a layout of $G$ is already given, and regard its $P$-stress $P_G$
as a function of $\ell$:
\begin{equation*}
    P_G(\ell) =
    \sum_{u < v \in V} w_{u v}
      \left( \left( \ell p_{u v} - b(u,v) \right)^{+} \right)^{2} +
    \sum_{(u,v) \in E} \ell^{-2}
      \left( \left( b(u,v) - \ell \right)^{+} \right)^{2}
\end{equation*}
where $b(u,v)$ is the Euclidean distance between the boundaries of nodes $u$
and $v$ along the straight line connecting their centres,
$p_{u v}$ the number of edges on the shortest path between nodes $u$ and $v$,
$w_{u v} = (\ell p_{u v})^{-2}$, and
$(z)^+=\max(z,0)$.
We wish to compute the value of $\ell$ that minimises $P_G$.

We begin by rewriting the $P$-stress as:
\begin{equation*}
    P_G(\ell) = 
    \sum_{(u,v) \in D} w_{u v}
      \left( \left( \ell p_{u v} - b(u,v) \right)^{+} \right)^{2} +
    \sum_{(u,v) \in E} \ell^{-2}
      \left( \ell - b(u,v) \right)^{2},
\end{equation*}
where $D = \{(u,v) : u < v \wedge (u,v) \not \in E \wedge (v,u) \not \in E \}$.
In other words, $D$ is simply the set of all ordered pairs written in ascending
order, in which the nodes are \emph{not} connected by an edge.

For each $(u,v) \in D$, define $\ell_{uv} = b(u,v)/p_{uv}$, and
\[
    h_{uv}(\ell) = \left\lbrace\begin{array}{cl}
        \left( 1 - \frac{\ell_{uv}}{\ell} \right)^2 & \ell \in [\ell_{uv},+\infty) \\
        0 & \ell \in (0, \ell_{uv}].
    \end{array}\right.
\]
Then $h_{uv}$ is in fact differentiable over $(0,+\infty)$, with
\[
    h_{uv}'(\ell) = \left\lbrace\begin{array}{cl}
        \frac{2\ell_{uv}}{\ell^2} \left( 1 - \frac{\ell_{uv}}{\ell} \right) & \ell \in [\ell_{uv},+\infty) \\
        0 & \ell \in (0, \ell_{uv}],
    \end{array}\right.
\]
and we have
\[
    P_G(\ell) =
    \sum_{(u,v) \in D} h_{uv}(\ell) +
    \sum_{(u,v) \in E} \left( 1 - \frac{b(u,v)}{\ell} \right)^2
\]
and
\[
    P_G'(\ell) =
    \sum_{(u,v) \in D} h_{uv}'(\ell) +
    \sum_{(u,v) \in E} \frac{2b(u,v)}{\ell^2}\left( 1 - \frac{b(u,v)}{\ell} \right).
\]

Now let $\langle \ell_1, \ell_2, \ldots, \ell_\nu\rangle$ be the list of all
$\ell_{uv}$ for $(u,v) \in D$, written in non-decreasing order (some values may appear more than once).
Since there may be repeated values among the $\ell_i$, let $\langle m_1, m_2, \ldots, m_\mu \rangle$
be the list of all \emph{distinct} values of the $\ell_i$, written in strictly ascending order.
For each $1 \leq j \leq \mu$, let $A_j = \{i : \ell_i <= m_j \}$, and
let $I_j = [m_j, m_{j+1}]$.
Restricting to the interval $I_j$ and substituting a new variable $\lambda_j$, we have
\[
    \left.P_G'(\lambda_j)\right|_{I_j} =
    \sum_{i \in A_j} \frac{2\ell_i}{\lambda_j^2}\left( 1 - \frac{\ell_i}{\lambda_j} \right) +
    \sum_{(u,v) \in E} \frac{2b(u,v)}{\lambda_j^2}\left( 1 - \frac{b(u,v)}{\lambda_j} \right),
\]
and setting this derivative equal to zero and solving for $\lambda_j$, we find
\[
    \lambda_j =
    \frac{ \sum_{(u,v) \in E} b(u,v)^2 + \sum_{i \in A_j} \ell_i^2 }
    { \sum_{(u,v) \in E} b(u,v) + \sum_{i \in A_j} \ell_i}.
\]
This is simply the contraharmonic mean
over $B \cup A_j$, where
$B = \langle b(u,v) : (u,v) \in E \rangle$; that is, the weighted mean in which the weights equal the values.
For $\lambda_j$ to be an actual critical point of the function $P_G$ however, it
must satisfy the assumption that it lies in the restricted interval $I_j$.
Thus the ideal edge length $\bar\ell$ is found to be
\[
    \bar\ell = \argmin_{\{\lambda_j : \lambda_j \in I_j\}} P_G(\lambda_j).
\]



\end{document}